\documentclass[referee]{raa}            

\usepackage{graphicx,times}             
\usepackage{natbib}
\usepackage{amssymb,amsmath}
\usepackage{booktabs}
\bibpunct{(}{)}{;}{a}{}{,}

\usepackage[pagebackref=true]{hyperref}

\begin{document}

  \title{Simulation and Data Processing of Beamforming Experiments with Four 21CMA Stations}


   \volnopage{Vol.0 (20xx) No.0, 000--000}      
   \setcounter{page}{1}          

  \author{Feiyu Zhao
   \inst{1,3}
    \and Quan Guo
   \inst{2} \thanks{E-mail: guoquan@shao.ac.cn}
   \and Junhua Gu
    \inst{4}  \thanks{E-mail: jhgu@nao.cas.cn}
   \and Qian Zheng
   \inst{2} \thanks{E-mail: qzheng@shao.ac.cn}
   \and Yan Huang
   \inst{4}
   \and Yun Yu
   \inst{2, 3}
   }

\institute{
   Shanghai Astronomical Observatory, Chinese Academy of Sciences, 80 Nandan Road, Shanghai 200030, China
   \and
   State Key Laboratory of Radio Astronomy and Technology, 
   Shanghai Astronomical Observatory, Chinese Academy of Sciences, 80 Nandan Road, Shanghai 200030, China; 
   \and
   University of Chinese Academy of Sciences, No.1 Yanqihu East Road, Beijing 101408, China\\
   \and
    State Key Laboratory of Radio Astronomy and Technology, National Astronomical Observatories, Chinese Academy of Sciences, 20A Datun Road, Beijing 100101, China }
    
\vs\no
   {\small Received 20xx month day; accepted 20xx month day}

\abstract{We present an end-to-end simulation and data-processing framework for digital beamforming experiments conducted with four stations of the 21Centimeter Array (21CMA). Motivated by the need to characterize instrumental systematics—such as those arising from station-level digital beam synthesis and two-stage channelization—and to validate the data-processing pipeline framework for a future upgraded 21CMA with beamforming capability across all stations, we simulate interferometric visibilities using realistic four-station layouts with radio interferometer simulation software \textit{OSKAR}. Two representative pointings are considered: a bright, complex Cassiopeia~A field and a near–north celestial pole (NCP) calibration field. The sky model combines cataloged point sources with a diffuse Galactic component from the Global Sky Model (GSM), and frequency-dependent thermal noise is injected. We further quantify the imprint of two-stage channelization by comparing an ideal beamformer with a coarse-channel phase approximation, demonstrating that off-axis sources exhibit a characteristic piecewise-linear spectral modulation across coarse-channel boundaries. A data-processing pipeline, including Radio Frequency Interference (RFI) mitigation, calibration, imaging, and mosaicking steps consistent with current low-frequency radio astronomy practice, is constructed. The resulting synthetic images and background root-mean-square (RMS) noise measurements demonstrate the feasibility of adapting established 21CMA calibration and imaging strategies to digital beamforming modes, and provide a framework that can be further developed for beam-aware processing in future full-scale 21CMA beamforming observations.
}

   \authorrunning{F.Zhao et al.}            
   \titlerunning{Simulation and Data Processing of Beamforming Experiments with Four 21CMA Stations}  

   \maketitle

%
%
\section{Introduction}           
\label{sect:intro}

The 21CMA is a ground-based radio interferometer located in the Ulastai Valley
of western China. Its array consists of 10,287 log-periodic antennas distributed along two perpendicular
arms in the east-west (E-W) and north-south (N-S) directions. Operating within a frequency range of $50\sim
200$ MHz, the instrument lacks beam-steering capability and is fixed to observe the North Celestial Pole, enabling continuous 24-hour monitoring of the same sky region. This design is optimized for deep integrations aimed at detecting the 
redshifted 21 cm signal from the Epoch of Reionization (EoR) or Cosmic Dawn (CD) \citep{2016ApJ...832..190Z,2022RAA....22a5012Z,2006PhR...433..181F,2012RPPh...75h6901P}.

The implementation of digital beamforming (DBF) on low-frequency aperture arrays is a key technological requirement for next-generation radio telescopes such as the Square Kilometre Array Low-Frequency Aperture Array (SKA LFAA) \citep{dewdney2016ska1,acedo2020ska}. In digital beamforming, signals from individual antenna elements are digitized and combined with frequency-dependent complex weights to synthesize one or more directive beams, enabling flexible beam steering and simultaneous multi-beam observations. To keep pace with recent advances in digital aperture-array technology and expand the scientific capabilities of the 21CMA, including low-frequency pulsar observations, we carry out a dedicated beamforming experiment using four of its stations. Similar station-based digital beamforming and two-stage channelization architectures have been widely adopted in modern low-frequency aperture arrays. In such systems, two-stage channelization typically consists of an initial coarse polyphase filter bank that partitions the wideband signal into subbands, followed by a second-stage fine channelization that provides the spectral resolution required for correlation, beamforming, and downstream analysis. This two-stage design reduces the downstream data rate and computational cost by allowing beamforming/correlation to operate on narrower subbands, while still delivering the fine spectral resolution needed for calibration and science analysis. For SKA LFAA, the reference signal-processing chain and station-level implementation have been documented in system design reports and in detailed studies of firmware and end-to-end beamformer/correlator modeling \citep{dewdney2016ska1,acedo2020ska,comoretto2017signal,chiarucci2020end}. 

In parallel, pathfinder experiments such as the Low-Frequency Array (LOFAR) \citep{van2013lofar}, the Murchison Widefield Array (MWA) \citep{tingay2013mwa}, and the Hydrogen Epoch of Reionization Array (HERA) \citep{deboer2017hera} have developed mature calibration and imaging pipelines for low-frequency radio interferometric observations, where realistic simulations are routinely used to validate instrumental systematics and spectral structures before (and alongside) real observations \citep{jelic2008foreground,patil2017upper,barry2019fhd,line2025verifying,aguirre2022validation}. A number of open-source simulation tools, including OSKAR, pyuvsim, and WODEN, enable visibility-level forward modeling and have become standard components in these end-to-end tests \citep{mort2010oskar,lanman2019pyuvsim,line2022woden}. Related low-frequency arrays such as The Long Wavelength Array (LWA) \citep{Ellingson2009LWA}, its expanded configuration the Owens Valley Radio Observatory Long Wavelength Array (OVRO-LWA) \citep{Eastwood2018OVROLWA}, and the New Extension in Nançay Upgrading LOFAR (NenuFAR) \citep{Zarka2020NenuFAR} have also reported early 21 cm power-spectrum analyses, further motivating the need for well-characterized beamforming responses and controlled pipeline verification \citep{ellingson2013design,eastwood201921,munshi2024first}.

Recent progress on low-frequency arrays has also established a clear methodological context for our work. 
Beamformed observing modes are routinely used for high-time-resolution science (e.g., pulsars and fast transients), which directly benefit from flexible digital station beamforming \citep{stappers2011observing}. 
For EoR/CD studies, robust 21 cm power-spectrum pipelines have been developed and tested on real data, with detailed end-to-end methodologies and early results demonstrated for the MWA \citep{jacobs2016murchison,beardsley2016first}. 
A central challenge in these analyses is the spectral structure introduced by instrument chromaticity, which motivates delay-space diagnostics and careful treatment of wide-field foreground leakage \citep{parsons2012per,thyagarajan2015foregrounds}. 
Accordingly, modern power-spectrum estimators commonly adopt statistically grounded frameworks that explicitly model foregrounds and data covariance, enabling controlled validation of analysis choices and instrumental systematics \citep{liu2011method,dillon2015empirical}. 
Taken together, these developments highlight the need for a dedicated, end-to-end simulation and data-processing framework that explicitly captures the spectral and chromatic effects introduced by station-level digital beamforming, and that can be used to validate calibration and power-spectrum pipelines for the upgraded 21CMA system.

In the original 21CMA configuration, all 127 antennas within a station are combined through an analog delay-and-sum network, resulting in a single fixed station beam \citep{2016ApJ...832..190Z,2022RAA....22a5012Z}.
This analog beam is transmitted to the central facility via optical fiber, digitized, and processed through a standard FX (Fourier-transform and cross-multiply) correlator, followed by calibration, imaging, and power-spectrum estimation \citep{offringa2014wsclean}.
This architecture provides only one analog-formed beam per station and does not support dynamic pointing or per-antenna signal calibration \citep{2016ApJ...832..190Z,2022RAA....22a5012Z}. In the digital-beamforming upgrade, each antenna is digitized independently. The station backend applies a coarse polyphase filter bank (PFB) \citep[PFB,][]{harris2011mathematical}, followed by frequency-domain beamforming to synthesize station beams in software. These digitally formed beams then undergo fine channelization before being sent to the FX correlator. The overall system follows the architecture described by \cite{acedo2020ska}. This architecture supports multiple simultaneous beams and dynamic pointing; however, it also introduces additional frequency-domain structures, for example, those associated with the two-stage channelization process, which will be discussed later.

To quantify these instrumental effects—particularly those associated with digital beamforming and channelization—and to validate the data-processing pipeline for the 21CMA beamforming observing mode, a dedicated simulation framework capable of generating synthetic observations for a four-station digital beamforming system is required. Such a framework enables realistic modeling of instrumental systematics and provides a controlled environment for developing and validating the corresponding data-processing pipeline. In this work, we develop a dedicated, end-to-end simulation framework for four-station digital beamforming observations with the 21CMA, explicitly incorporating the systematics introduced by digital beamforming and frequency channelization. Using this framework, we generate realistic synthetic observations, construct the associated data-processing pipeline, and demonstrate each stage of calibration and imaging using physically motivated sky models \citep{de2008model,gorski2005healpix,intema2017gmrt,rengelink1997westerbork}.
This framework provides a unified pathway from synthetic visibilities to final, science-ready images and serves as a foundation for future improvements in the data processing of beamforming observations with the upgraded 21CMA.

\section{Beamforming Experiments with four 21CMA Stations}
\subsection{The system of Beamforming Experiment}
The 21CMA system provides a high frequency resolution of 24.4 kHz. Data from each frequency channel are integrated over 3.56-second intervals before being recorded to a disk array. The baseline vector between antennas is described by the components $(u, v, w)$ in a right-handed coordinate system, where $u$ and $v$ lie in a plane perpendicular to the phase reference direction (with $u$ pointing east and $v$ pointing north), and $w$ is aligned along the line of sight toward the phase reference position. For E-W baselines of 21CMA, which have no component parallel to the Earth's rotation axis, the $w$-component of the baseline vector can be neglected when the synthesized field of view is not excessively large. This configuration simplifies the imaging algorithm by avoiding the complications introduced by the non-coplanar baselines, a problem known as the $w$-term. The longest E-W baseline  of the full 21CMA is 2780 meters, which yields an angular resolution of approximately $2^\prime$ at 200 MHz for resolving source structures.

The overall implementation of our beamforming experiment follows the SKA LFAA signal processing chain (e.g., \citealt{comoretto2017signal,chiarucci2020end}).
Our experiment reuses the existing 21CMA infrastructure while integrating a new digital acquisition and beamforming system for four stations (E13, E03, W02 and W09). Each station retains its original array of 127 log-periodic antennas, whose signals pass through the upgraded radio-frequency (RF) chain consisting of low-noise amplifiers, bandpass filters covering 50--200 MHz, and adjustable attenuators for gain equalization. 

Digitization and channelization are carried out in the Station Processing Unit (SPU), where time-domain signals are converted to coarse spectral channels using an oversampled PFB. Digital beamforming is implemented in the frequency domain via complex weighting applied to each coarse channel. The beamformed outputs are transported over 10 Gbps optical links to the central machine room. A second-stage fine channelizer, implemented on graphics processing unit (GPU) hardware, produces high-resolution spectral streams for correlation or beam-power analysis.

Accurate station synchronization is achieved using a White Rabbit (WR) \citep[WR,][]{Moreira2009WhiteRabbit} network, which distributes both the 10 MHz frequency reference and the Pulse Per Second (PPS) timing signal to all stations with sub-nanosecond precision. Monitoring, control, and metadata acquisition are performed via a dedicated 1 Gbps Ethernet network. This architecture allows real-time beam steering, dual-beam operation, and flexible switching between standard visibility mode and high-time-resolution modes for pulsar or transient observations.

The end-to-end system integrates the analog, digital, timing, and data-transport subsystems into a unified operational pipeline. The upgraded RF frontend ensures stable gain and phase characteristics for all antennas in the four participating stations. After digitization, coarse PFB channelization and frequency-domain beamforming are performed on field-programmable gate array (FPGA)-based Digital Beamforming Units (DBUs), which apply phase delays computed from the desired pointing direction. The beamformed coarse-channel data packets are transferred via optical fibers to the central machine room, where GPU-based fine channelization and correlation are executed. 

The architecture supports continuous beam tracking of strong calibration sources, rapid station reconfiguration, and high-rate data recording. Details of the system design and implementation will be presented in Gu et al. (2026, in preparation). Four 21CMA stations (E13, E03, W02, and W09) were selected from the existing north–south and east–west arms of the array, and all are equipped with the new SPUs, DBUs, and WR timing receivers. The longest baseline in this four-station sub-array is 1480 m (E13--W09). This selection ensures a mix of short and moderate baselines, allowing us to evaluate fringe stability, inter-station phase coherence, and the spectral behavior of the digitally formed beams under a variety of source directions. 

\section{Simulation}
\label{sect:Obs}

\begin{figure*}
  \centering
  \begin{subfigure}[t]{0.48\textwidth}
    \centering
    \includegraphics[width=\textwidth]{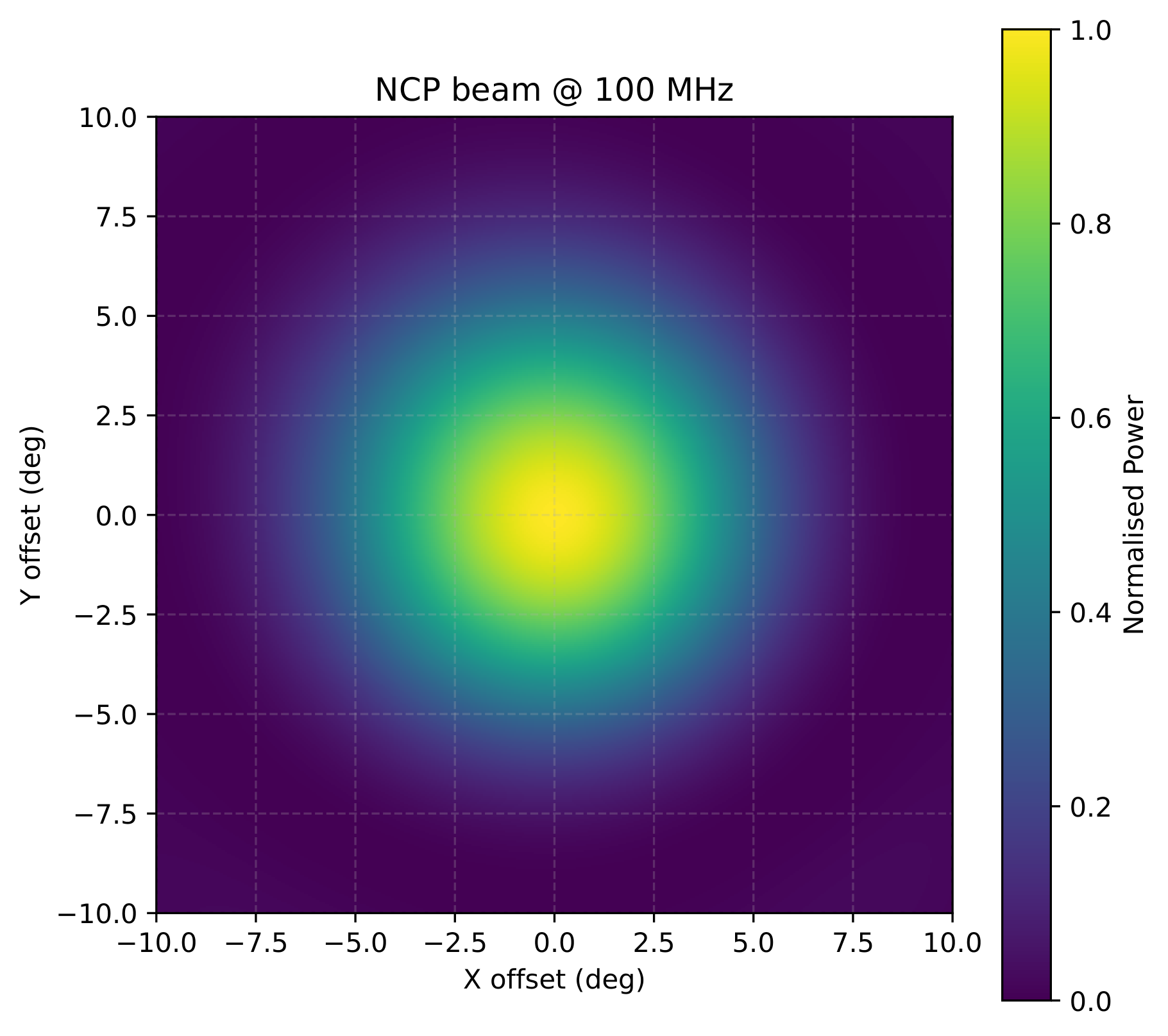}
    \caption{NCP pointing.}
    \label{fig:beam_ncp}
  \end{subfigure}\hfill
  \begin{subfigure}[t]{0.48\textwidth}
    \centering
    \includegraphics[width=\textwidth]{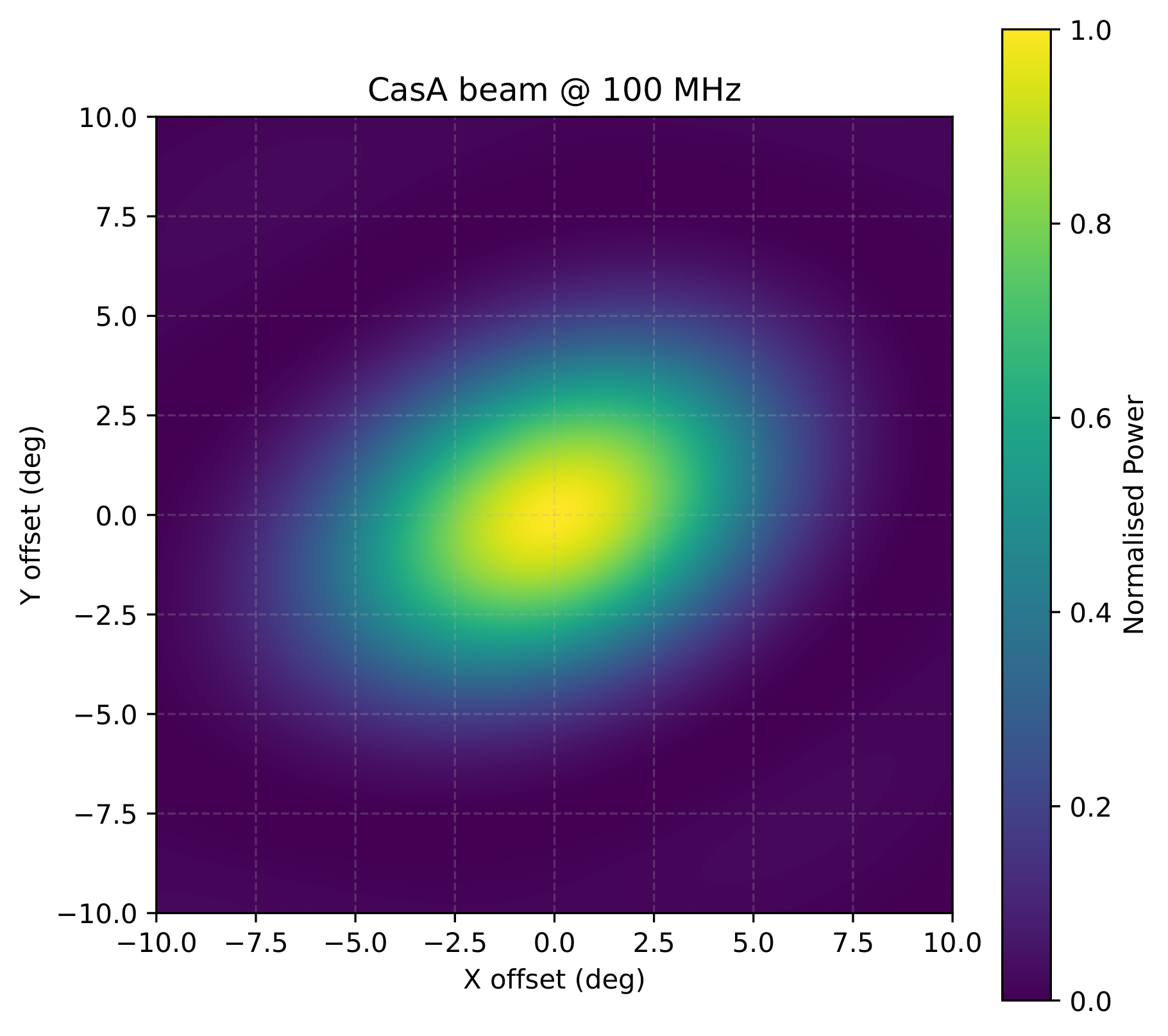}
    \caption{Cas~A pointing.}
    \label{fig:beam_casa}
  \end{subfigure}
  \caption{Simulated 21CMA station beam patterns at 100\,MHz for the two target fields used in this work. The beams are computed for the four–station sub–array (E13, E03, W02 and W09) and are normalized to unity at the respective phase centers.}
  \label{fig:21cma_beams}
\end{figure*}

We simulated interferometric observations of the 21CMA using the OSKAR software\footnote{\url{https://github.com/OxfordSKA/OSKAR}}. We selected four stations consisting of stations E13, E03, W02, and W09, which lie along the predominantly east-west arms of 21CMA. Each station was modeled as a beam-formed aperture array based on the station layouts, with identical isotropic, unpolarized elements so that the station beam is determined purely by the array factor.
The corresponding station beams at 100 MHz are shown in Figure~\ref{fig:21cma_beams}. 

\begin{figure*}
  \centering
  \begin{subfigure}[t]{0.48\textwidth}
    \centering
    \includegraphics[width=\textwidth]{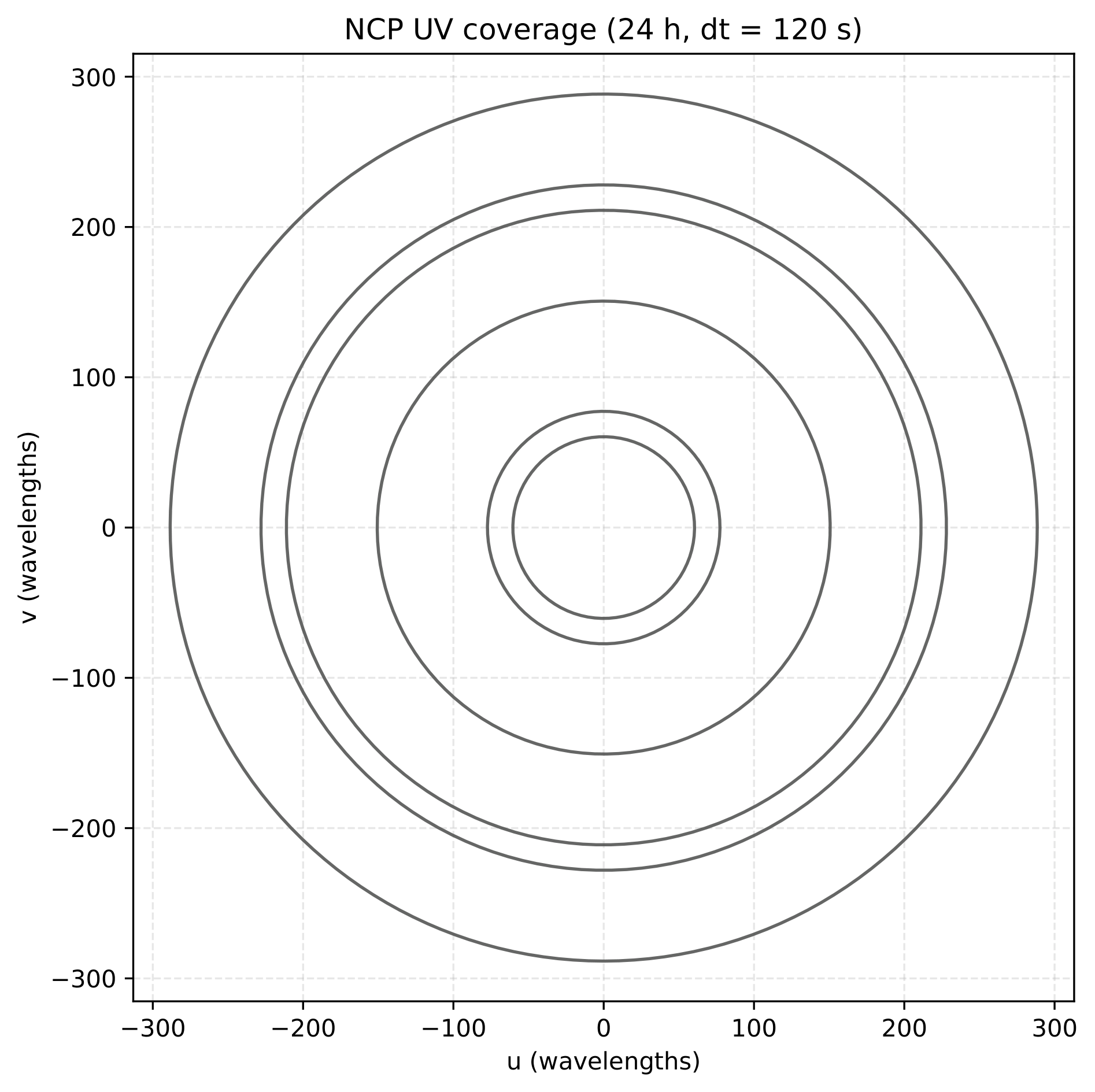}
    \caption{NCP $uv$ coverage.}
    \label{fig:uv_ncp}
  \end{subfigure}\hfill
  \begin{subfigure}[t]{0.48\textwidth}
    \centering
    \includegraphics[width=\textwidth]{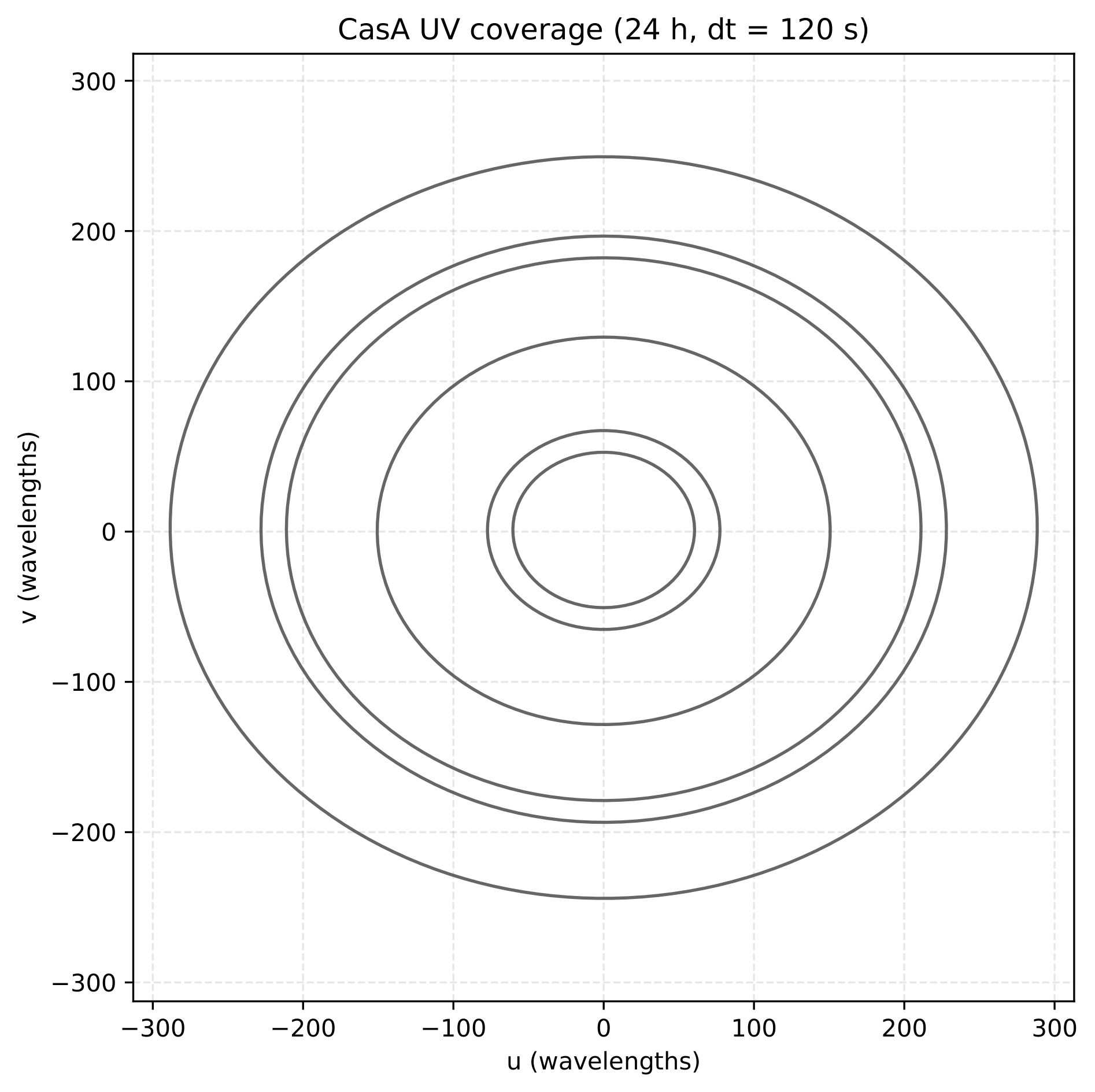}
    \caption{Cas~A $uv$ coverage.}
    \label{fig:uv_casa}
  \end{subfigure}
  \caption{Simulated 24-hour $uv$ coverages at 100\,MHz for the NCP and Cassiopeia~A fields, using the four 21CMA stations E13, E03, W02 and W09. Baselines are plotted in units of observing wavelength.}
  \label{fig:21cma_uvcov}
\end{figure*}

For each target field, we generated a continuous 24-hour tracking observation with regular time sampling. The associated 24-hour $uv$ coverages are shown in Figure~\ref{fig:21cma_uvcov}. Simulated visibilities cover a contiguous 50–200~MHz band with 1~MHz channel spacing, which provides sufficient spectral resolution to follow the frequency dependence of the primary beam and the sky emission. In the analysis presented here we focus on the central 100–150~MHz band, but all simulations were carried out over the full 50–200~MHz range.

We performed two simulations: a full end-to-end 24-hour visibility simulation with 1 MHz frequency sampling for imaging and pipeline validation, and a separate, minimal 5 MHz test focusing on the two-stage channelization response using a single point source and the coarse-channel phase approximation.

\begin{figure}
  \centering
  \begin{subfigure}{0.48\columnwidth}
    \centering
    \includegraphics[width=\linewidth]{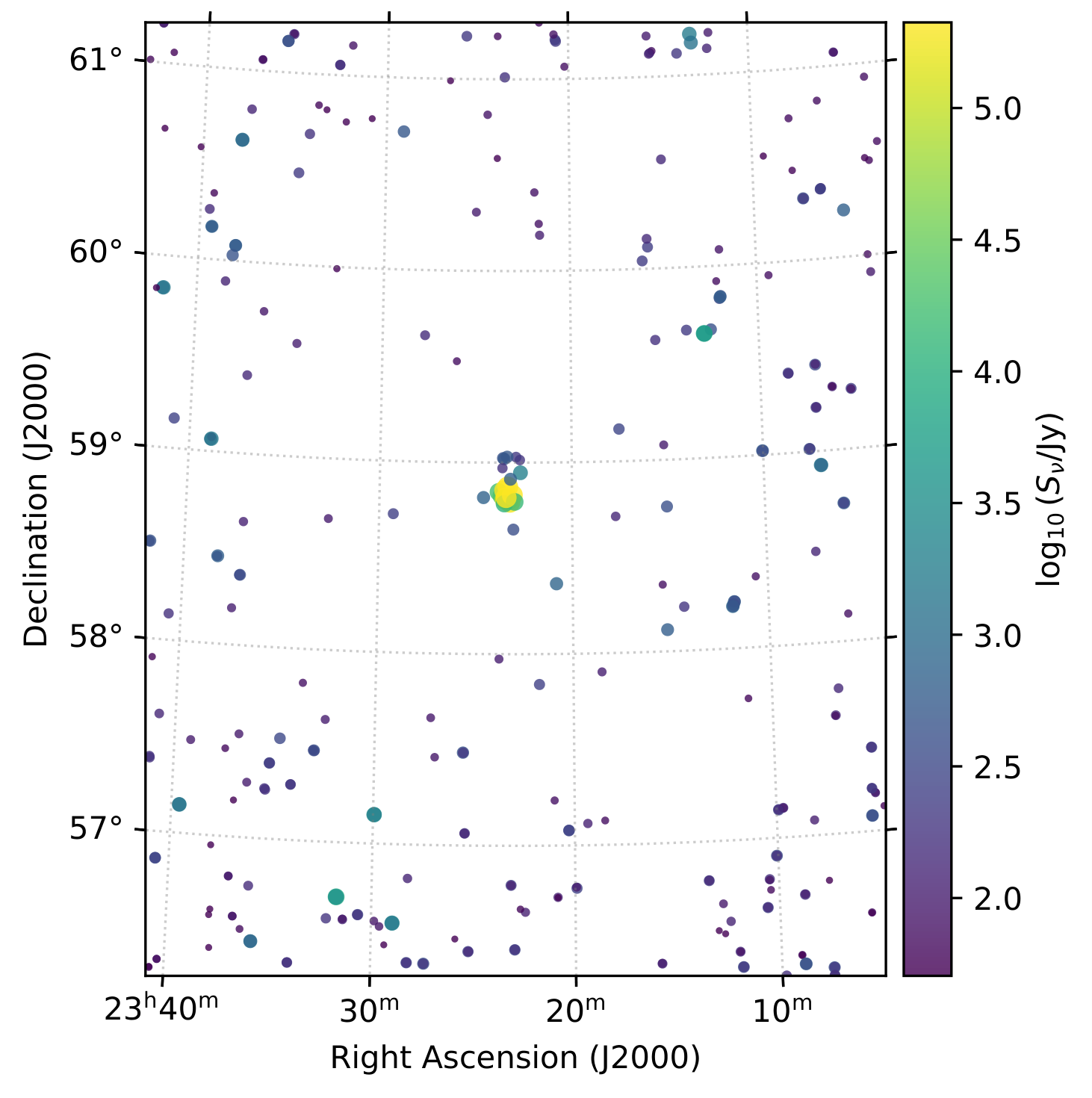}
    \caption{Cassiopeia~A field.}
    \label{fig:sky_casa}
  \end{subfigure}
  \hfill
  \begin{subfigure}{0.48\columnwidth}
    \centering
    \includegraphics[width=\linewidth]{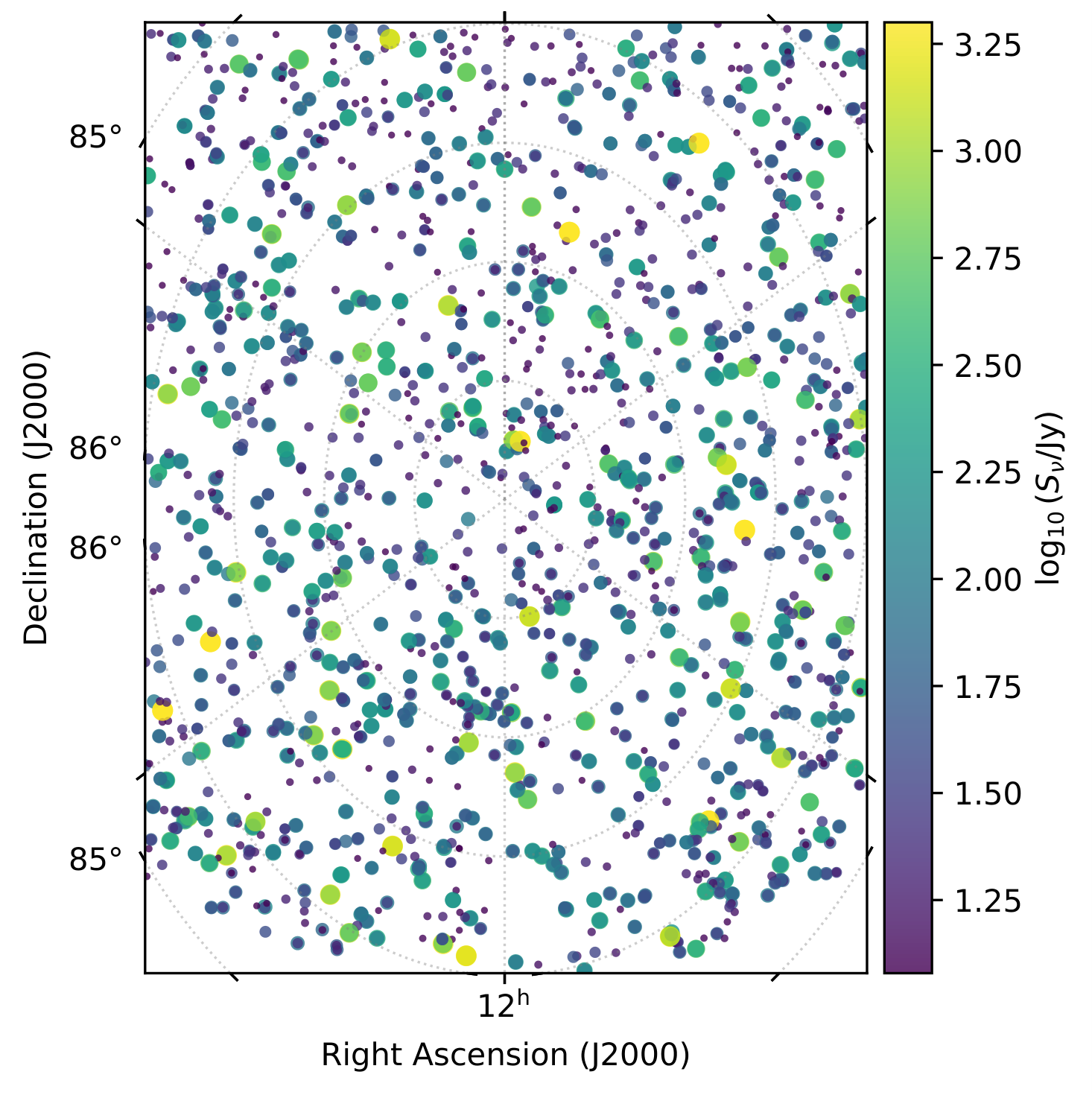}
    \caption{Near–NCP field.}
    \label{fig:sky_ncp}
  \end{subfigure}
  \caption{Input point-source sky models used for the simulated 21CMA observations. The panels show the distribution of cataloged radio sources in equatorial coordinates (J2000) for the Cassiopeia~A field (left; 5$^\circ$ field of view) and the near--north celestial pole calibration field (right; 8$^\circ$ field of view). The source flux is encoded by color. A diffuse Galactic component from a GSM realization and thermal noise are also included in the simulations but are not shown here.}
  \label{fig:sky_models}
\end{figure}

Two pointings were simulated with this setup: one centered on Cassiopeia~A and one on the north celestial pole. These represent, respectively, an extremely bright source field with complex surrounding structure and a continuously visible field corresponding to the standard 21CMA observing mode. The corresponding point-source sky models used in these simulations are shown in Figure~\ref{fig:sky_models}, where each point marks a cataloged source color-coded by its logarithmic flux density. Within the $5^\circ$-diameter imaging field centered on Cassiopeia~A, the sky model contains 309 sources with flux densities in the range $4.1\times10^{1}$--$7.9\times10^{5}$~Jy (median $1.6\times10^{2}$~Jy). In the $8^\circ$ field around the near--NCP calibration region, 2136 sources fall within the imaged area, with flux densities spanning $7$--$3.7\times10^{4}$~Jy (median $5.7\times10^{1}$~Jy). Each point source is assigned a simple power-law spectrum,
\begin{equation}
  S_\nu = S_{\nu_0}\,\left(\frac{\nu}{\nu_0}\right)^{\alpha_{\rm ps}},
\end{equation}
with spectral index $\alpha_{\rm ps} = -0.7$ and reference frequency $\nu_0$ equal to the cataloged flux-density frequency (typically around 150~MHz for low-frequency surveys such as the TIFR GMRT Sky Survey (TGSS) and the Westerbork Northern Sky Survey (WENSS), \citealt{rengelink1997westerbork},\citealt{intema2017gmrt}).

In addition to the discrete source catalogs, we included a diffuse Galactic foreground component derived from a full-sky GSM, \citep[e.g.][]{de2008model} evaluated at a reference frequency of $\nu_0 = 75$~MHz. The GSM map is provided in Hierarchical Equal Area isoLatitude Pixelization (HEALPix) format and gives the brightness temperature $T_{\rm b}(\nu_0,\hat{\boldsymbol{n}})$ as a function of sky direction $\hat{\boldsymbol{n}}$. For each HEALPix pixel we convert $T_{\rm b}$ to a flux density at $\nu_0$ using the Rayleigh--Jeans relation
\begin{equation}
  I_\nu(\nu_0,\hat{\boldsymbol{n}}) = \frac{2 k_{\rm B} T_{\rm b}(\nu_0,\hat{\boldsymbol{n}})}{\lambda_0^2}, \qquad
  S_\nu(\nu_0,\hat{\boldsymbol{n}}) = I_\nu(\nu_0,\hat{\boldsymbol{n}})\,\Omega_{\rm pix},
\end{equation}
where $\lambda_0 = c / \nu_0$ is the wavelength and $\Omega_{\rm pix}$ is the solid angle of a HEALPix pixel. Each pixel is then represented in OSKAR as an extended source located at the pixel center. The frequency dependence of this diffuse component is modeled as a power law in brightness temperature,
\begin{equation}
  T_{\rm sky}(\nu) = T_{\rm sky}(\nu_0)\,\left(\frac{\nu}{\nu_0}\right)^{\beta_{\rm diff}},
\end{equation}
with spectral index $\beta_{\rm diff} \simeq -2.5$, which corresponds to a flux-density spectral index
\begin{equation}
  S_\nu \propto \nu^{\alpha_{\rm diff}}, \qquad \alpha_{\rm diff} = 2 + \beta_{\rm diff} \simeq -0.5.
\end{equation}
In the simulations we therefore assign a fixed spectral index $\alpha_{\rm diff} = -0.5$ to all GSM-derived components, referenced to $\nu_0 = 75$~MHz. In practice we restrict the number of GSM pixels used in OSKAR by grouping nearby pixels, so that the total number of diffuse components remains computationally manageable while still tracing the large-scale Galactic emission.

Thermal noise was added directly to the simulated visibilities using the standard radiometer equation. 
For each frequency channel $\nu$ we model the sky contribution to the system temperature as a power law,
\begin{equation}
  T_{\rm sky}(\nu)=A_{\rm sky}\left(\frac{\nu}{\nu_{\rm ref}}\right)^{\alpha_T},
\end{equation}
with $A_{\rm sky}=60~{\rm K}$ at $\nu_{\rm ref}=300$~MHz and $\alpha_T=-2.55$. 
We adopt a frequency-independent receiver contribution $T_{\rm tel}=50$~K, giving
\begin{equation}
  T_{\rm sys}(\nu)=T_{\rm tel}+T_{\rm sky}(\nu).
\end{equation}
The system-equivalent flux density (SEFD) is computed per station as (\citealt{thompson2017interferometry})
\begin{equation}
  {\rm SEFD}(\nu)=\frac{2k_{\rm B}T_{\rm sys}(\nu)}{\eta\,A_{\rm eff}}\times 10^{26}~{\rm Jy},
\end{equation}
where $A_{\rm eff}=216~{\rm m}^2$  is the effective area of a single 21CMA station, which remains approximately constant over our observing frequency range given the antenna design,  and $\eta = 0.5$ is the current efficiency factor for the 21CMA\citep{2016ApJ...832..190Z}. For a channel bandwidth of $\Delta\nu=1$~MHz and an integration time $\tau=60$~s,
The thermal-noise standard deviation of each visibility component is
\begin{equation}
  \sigma_{\rm N}(\nu)=\frac{{\rm SEFD}(\nu)}{\sqrt{2\,\Delta\nu\,\tau}}.
\end{equation}

We tabulate $\sigma_{\rm N}(\nu)$ over the observing band and provide it to OSKAR as the per-channel station noise RMS. 
OSKAR then injects uncorrelated Gaussian noise into the real and imaginary parts of each visibility sample for every baseline and time--frequency bin.

\begin{figure*}[t]
  \centering
  \includegraphics[width=\textwidth]{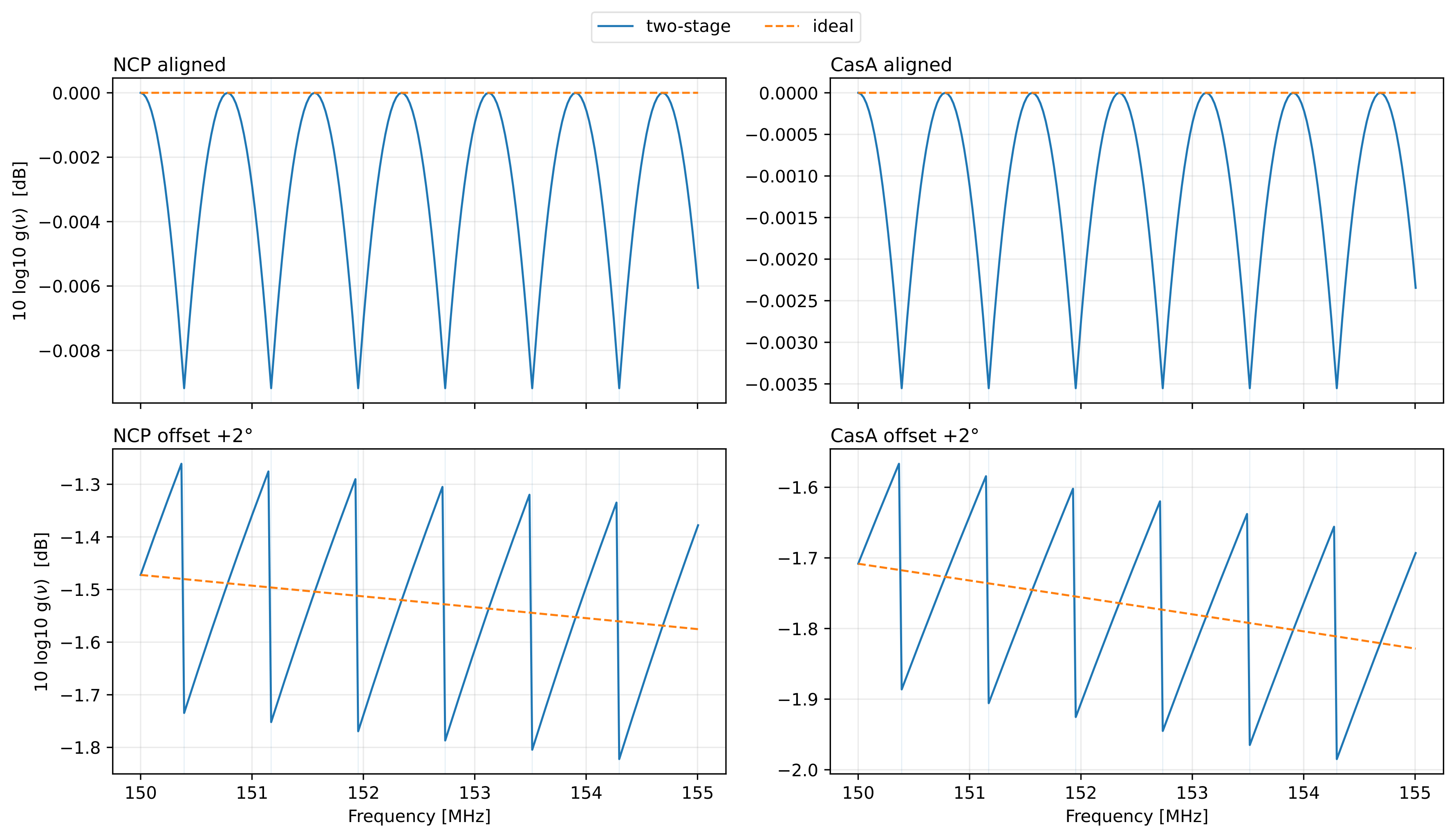}
  \caption{Two-stage channelization DBF imprint on the station power response for the NCP and Cas\,A pointings using the measured 127-antenna station layout. Curves show $10\log_{10}g(\nu)$ (dB) over a 5\,MHz band: dashed lines correspond to the ideal case ($\nu'=\nu$), and solid lines to the two-stage approximation where the beamforming phase is evaluated at a coarse-channel center frequency $\nu'$ that is held fixed within each coarse channel ($B=781.25$~kHz; faint vertical lines mark coarse-channel boundaries). Top: on-axis source. Bottom: source offset by $2^{\circ}$ towards zenith.}
  \label{fig:two_stage_channelization_g_2x2}
\end{figure*}

We additionally quantify the imprint of two-stage channelization DBF on the station response using a minimal setup (a single 1\,Jy point source at one representative time sample over a 5\,MHz band) for the NCP and Cas\,A pointings, evaluated with the measured 127-antenna station layout. In an ideal beamformer, the per-antenna phase is evaluated at the fine-channel frequency $\nu$, whereas in the two-stage approximation it is evaluated at a piecewise-constant coarse-channel center frequency $\nu'$ across each coarse channel. This is captured by the station voltage sum
\begin{equation}
\tilde U(\nu,\hat{\mathbf n};\nu',\hat{\mathbf n}_0)
= \frac{1}{N}\sum_{k=1}^{N}\exp\!\left[i\,\frac{2\pi}{c}\,\big(\nu\,\hat{\mathbf n}-\nu'\,\hat{\mathbf n}_0\big)\cdot\mathbf x_k\right],
\label{eq:two_stage_dbf}
\end{equation}
where $\mathbf x_k$ are the station element positions, $\hat{\mathbf n}_0$ is the pointing direction, and $\nu'$ is held fixed within each coarse channel (of width $B=781.25$~kHz in our setup). We define $g(\nu)\equiv|\tilde U(\nu,\hat{\mathbf n};\nu',\hat{\mathbf n}_0)|^2$ as the station power response. Figure~\ref{fig:two_stage_channelization_g_2x2} compares the ideal case ($\nu'=\nu$) with the two-stage approximation and shows that for an on-axis source the difference is small, whereas for a source offset by $2^{\circ}$ towards zenith the two-stage DBF introduces a characteristic piecewise-linear ``sawtooth'' modulation across frequency, superposed on the smooth chromatic trend of the ideal response.

\section{Data Processing}
\label{sect:data}

The data were processed using a data-processing strategy consistent with current practice for wide-field, multi-beam low-frequency radio interferometers such as the MWA. The previous calibration techniques of 21CMA datasets are also adopted in this pipeline (\citealt{2016ApJ...832..190Z}, \citealp{2022RAA....22a5012Z}). The beamforming observation mode exhibits station-based beam patterns that vary across the field of view, necessitating direction-dependent calibration and imaging as well as beam-aware calibration to accurately recover both compact and diffuse emission. Accordingly, the data are processed using a workflow adapted for station-level digital beamforming observations and consistent with current practice for wide-field, multi-beam low-frequency radio interferometers. Raw visibilities are first inspected and pre-processed to identify and excise radio-frequency interference, followed by appropriate time and frequency averaging. Initial direction-independent calibration is performed using bright calibrator sources to establish bandpass and complex gain solutions, which are subsequently refined through direction-dependent, beam-aware calibration to account for spatially varying station beam responses. Imaging is carried out using wide-field algorithms that properly handle non-coplanar baselines and frequency-dependent primary beams, and multiple synthesized beams are mosaicked to produce the final images, enabling a quantitative assessment of imaging fidelity and noise performance under the beamforming observing mode.

\subsection{RFI removal}

\begin{figure}
  \centering
  \includegraphics[width=0.85\columnwidth]{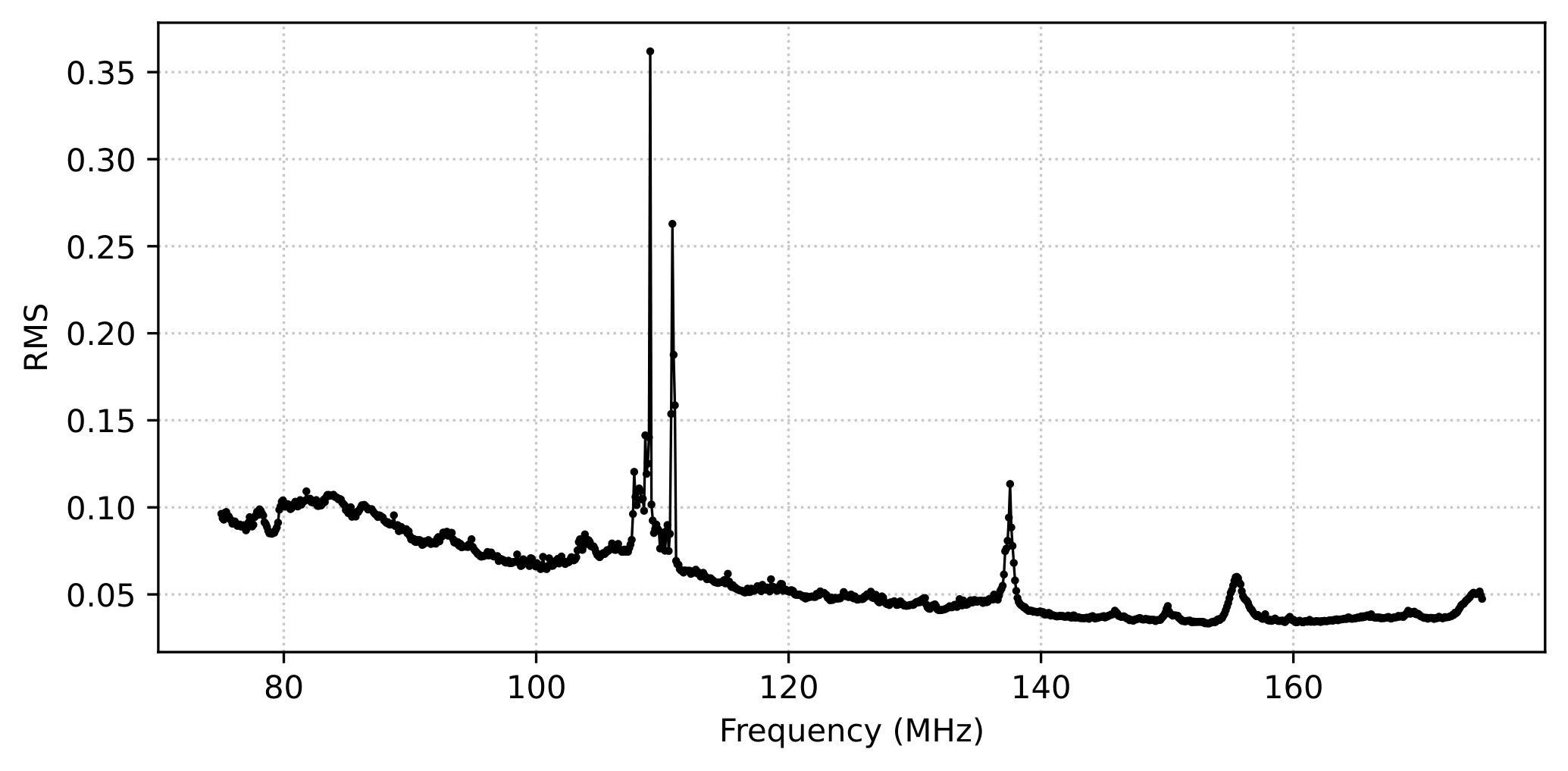}
  \caption{Per-channel RMS noise as a function of frequency for the 21CMA observations.}
  \label{fig:rfi_vs_freq}
\end{figure}

 RFI mitigation is a critical challenge in the reduction of low-frequency radio data. While the 21CMA site benefits from a relatively quiet radio environment, 21CMA observations can contain persistent RFI from several time-variable sources, including amplitude modulation (AM; $\sim$70 MHz) and scattered frequency modulation (FM; 88–108 MHz) radio broadcasts, civil aviation communications ($\sim$119 and $\sim$ 130 MHz), low-orbiting satellites ($\sim$137 MHz), and local train communications ($\sim$150 MHz) (\citealt{huang2016radio}). An initial flagging step identifies and removes these known contaminants directly from the visibility data. Figure~\ref{fig:rfi_vs_freq} presents the distribution of RFI as a function of frequency in the observed 21CMA dataset. In the current version of our simulations, no RFI is injected. Therefore, this flagging step is not applied in the simulated data processing. Here, we briefly demonstrate the RFI mitigation method as it would be applied to observational data. A more complete simulation framework that includes realistic RFI contamination will be implemented in future work.

Raw visibilities were first preprocessed to remove instrumental artifacts and RFI. We employed automated flagging based on statistical outlier detection in both time and frequency domains, supplemented by manual inspection of affected intervals. Autocorrelations and known persistent spectral contaminants were discarded. We subsequently apply a statistical algorithm for more comprehensive RFI excision. This process treats the real and imaginary components of the visibilities separately. For each component, data points deviating by more than 3$\sigma$ from a Gaussian distribution are flagged and excised. A visibility point is rejected if either its real or imaginary part exceeds this threshold. 

\subsection{Calibration}
The data were then amplitude-scaled and delay-corrected using a primary flux density calibrator observed in the same bandpass. Complex gain solutions were derived per antenna (or per station tile) and interpolated over the target observation.

Because the primary beams of beam-forming systems vary across the field, and ionospheric phase fluctuations are non-uniform, we adopted a direction-dependent calibration approach. A sky model constructed from existing catalogs and in-field bright sources was iteratively refined during calibration. Phase solutions were solved independently toward multiple directions across the field of view, allowing spatial modeling of ionospheric distortions. Where signal-to-noise permitted, we additionally solved for slow time variation in beam shape using a parameterized primary-beam model.

Our calibration strategy for the 21CMA experiments is sky-model based and follows previous analyses of the NCP region (\citealt{2016ApJ...832..190Z}, \citealp{2022RAA....22a5012Z}). In Zheng (2016), six bright sources near the NCP were used as primary calibrators, while Zhao (2022) constructed a $\sim$40--source sky model for flux and phase calibration as well as primary--beam correction. In this work, we consider two target fields. One field is centered on Cassiopeia~A at (RA, Dec) = (23$^\mathrm{h}$23$^\mathrm{m}$, +58$^\circ$48$^\prime$), where a dense sky model dominated by Cassiopeia~A and several other bright sources can be used to drive self-calibration. The other field is centered on the NCP, which is observable in the normal tracking mode of the 21CMA. The sky model for the NCP region can be adopted directly from the previous 21CMA works (\citealt{2016ApJ...832..190Z}, \citealp{2022RAA....22a5012Z}). To construct the sky model of the Cassiopeia~A field, we compiled radio sources within a $5^\circ$ radius centered on Cassiopeia~A using a combination of low-frequency radio surveys below 1 GHz. The resulting catalog includes each source's J2000 position, integrated flux density, and reference frequency, primarily drawn from TGSS~ADR1 (150 MHz) and WENSS (325 MHz). The input sky models used in these simulations shown in Figure~\ref{fig:sky_models} can be used for calibrations with future observation data reduction in these two sky regions.

\subsection{Imaging, primary beam correction and mosaicking }

\begin{figure*}[t]
  \centering

  \begin{subfigure}{0.48\textwidth}
    \centering
    \includegraphics[width=\linewidth]{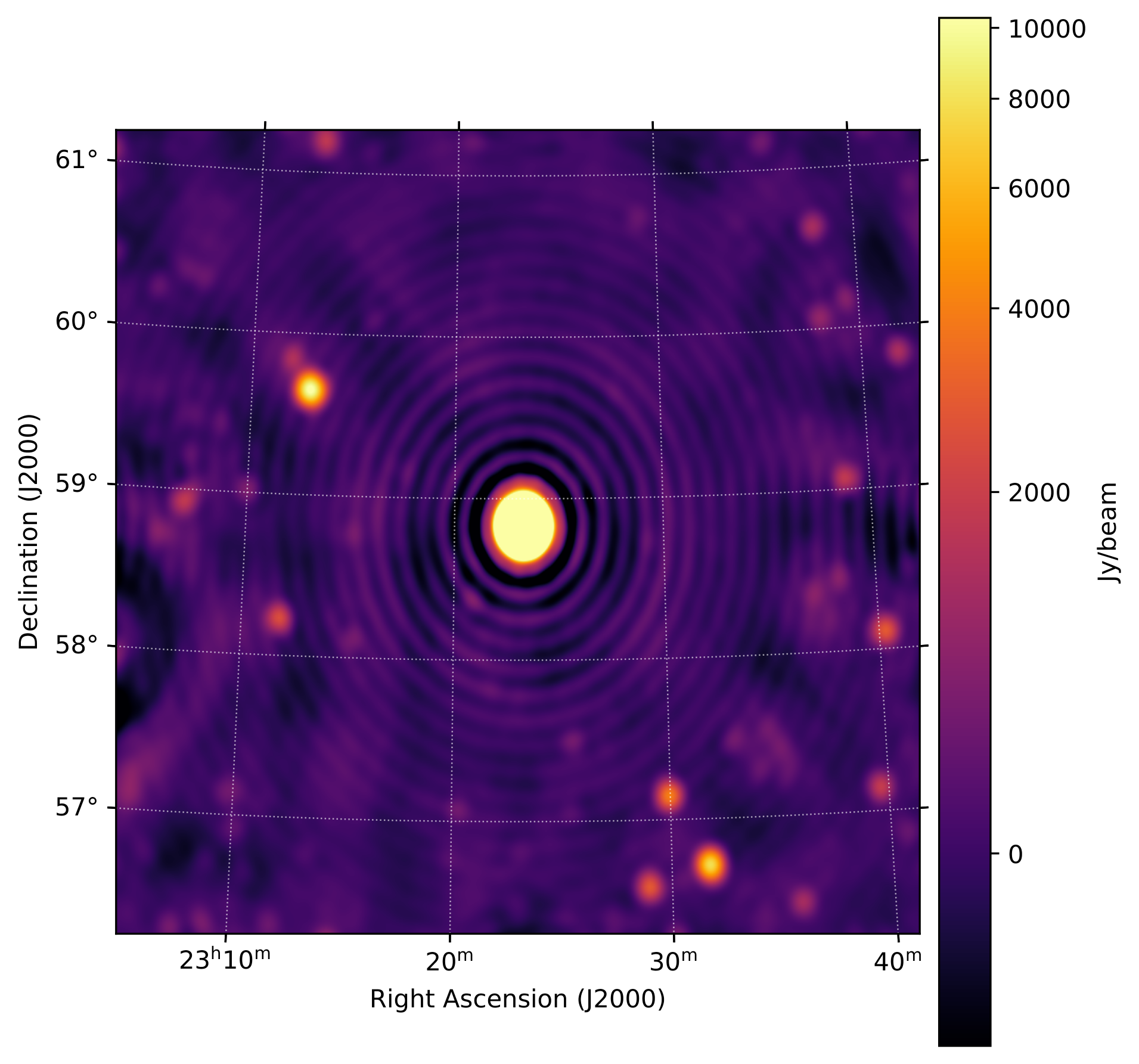}
    \caption{Cas~A: CLEAN image}
    \label{fig:casA_mfs_clean}
  \end{subfigure}
  \hfill
  \begin{subfigure}{0.48\textwidth}
    \centering
    \includegraphics[width=\linewidth]{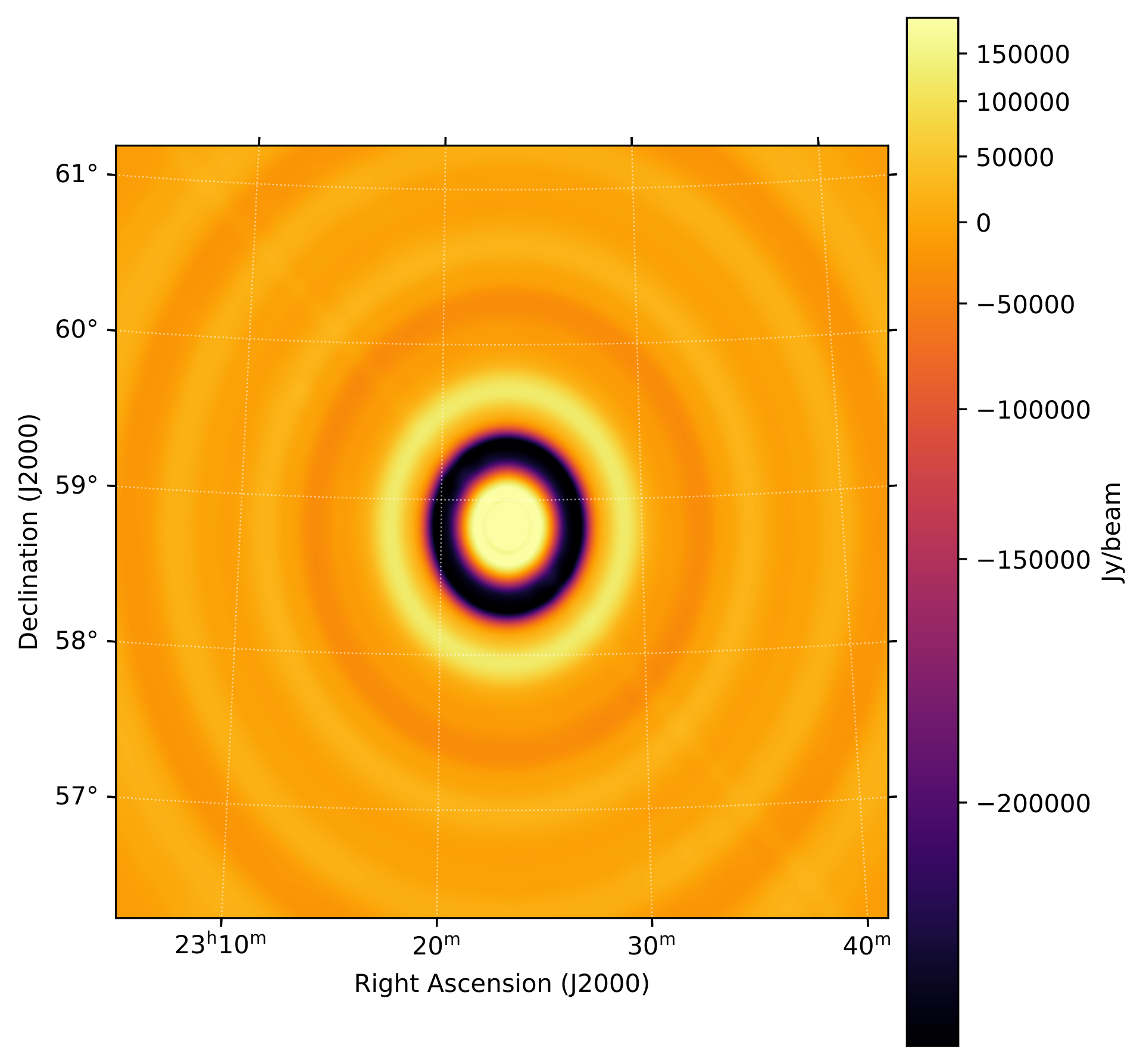}
    \caption{Cas~A: dirty image}
    \label{fig:casA_mfs_dirty}
  \end{subfigure}

  \vspace{0.6em}

  \begin{subfigure}{0.48\textwidth}
    \centering
    \includegraphics[width=\linewidth]{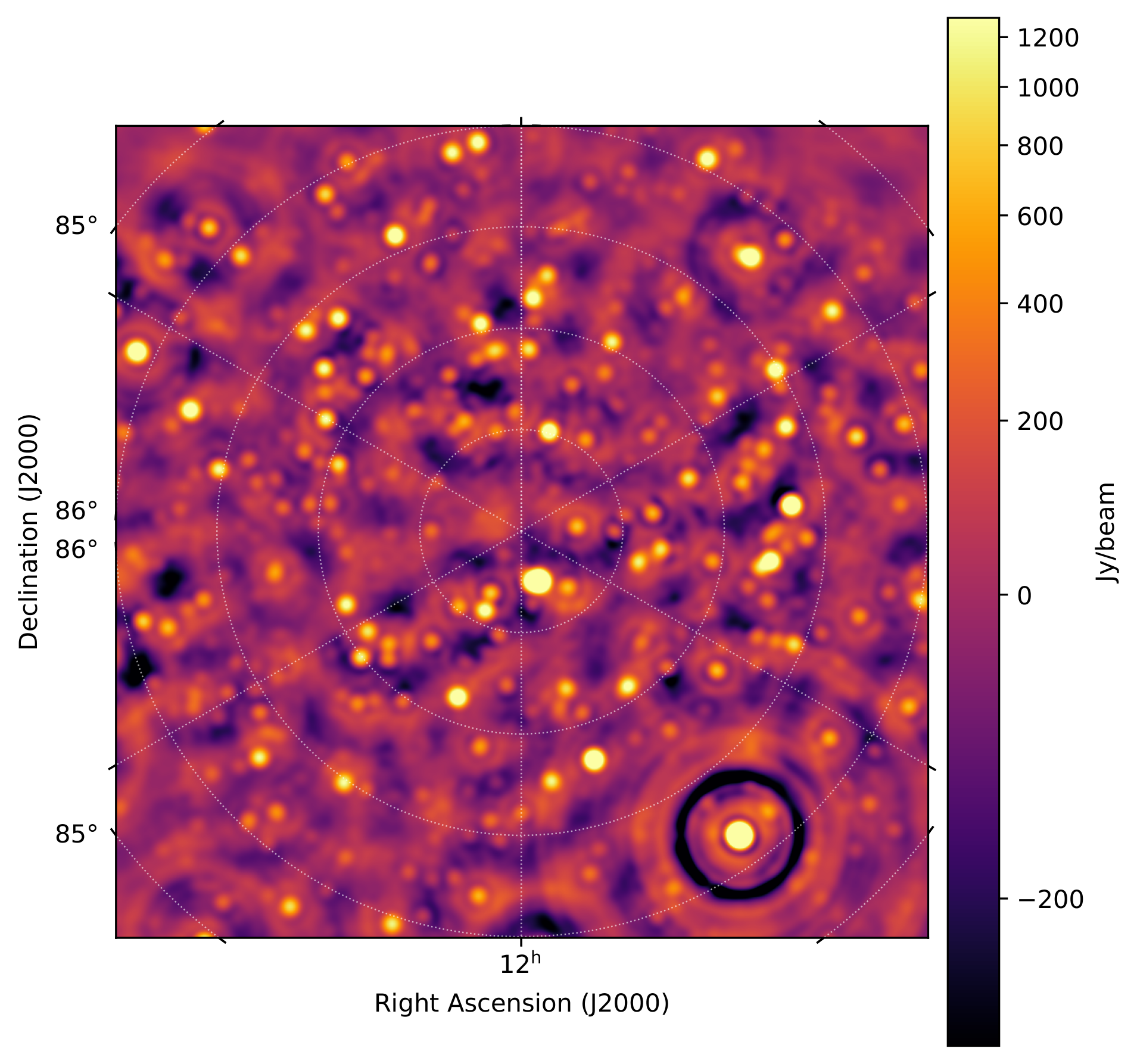}
    \caption{NCP: CLEAN image}
    \label{fig:ncp_mfs_clean}
  \end{subfigure}
  \hfill
  \begin{subfigure}{0.48\textwidth}
    \centering
    \includegraphics[width=\linewidth]{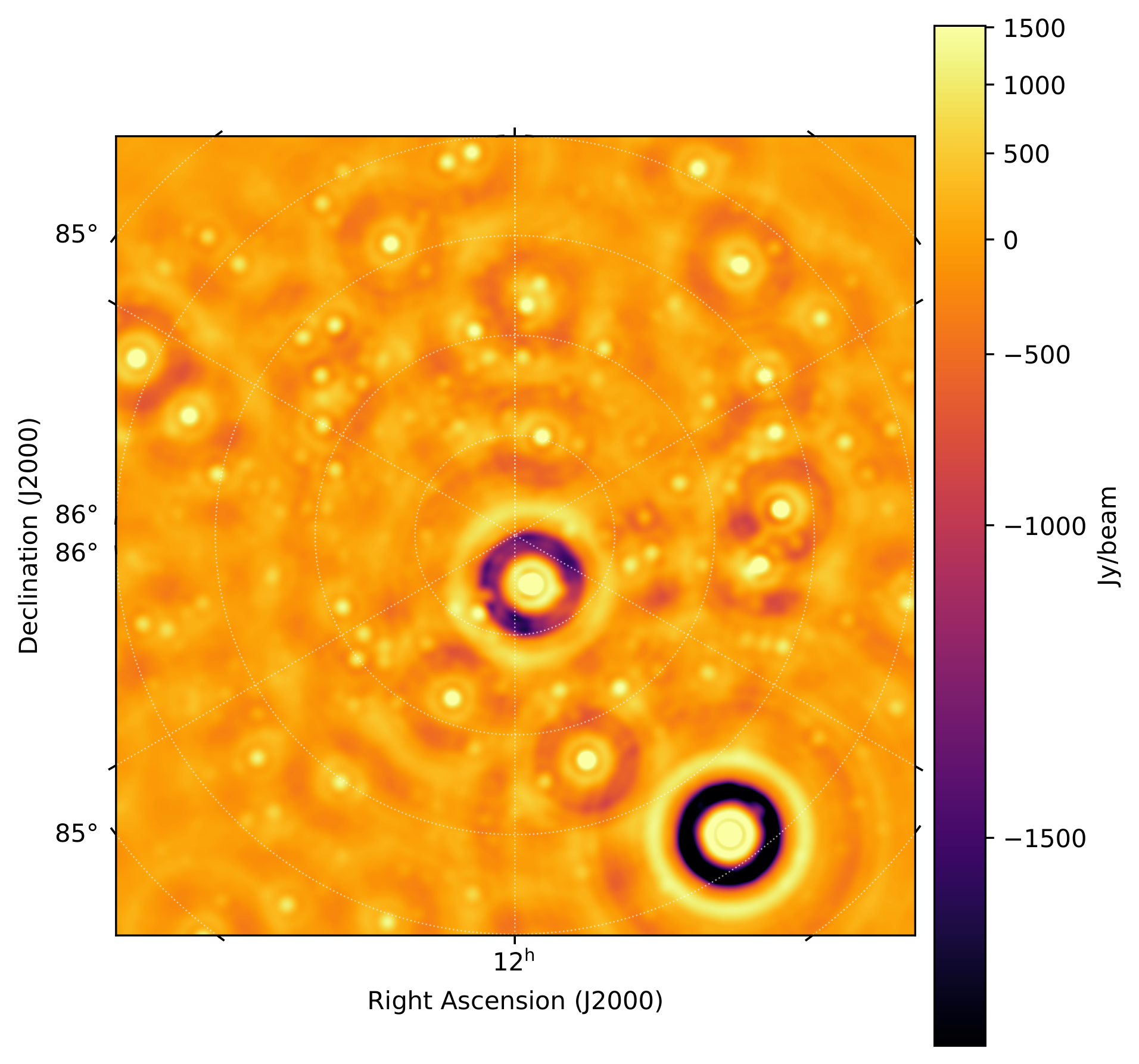}
    \caption{NCP: dirty image}
    \label{fig:ncp_mfs_dirty}
  \end{subfigure}

  \caption{Simulated 24-hour multi-frequency synthesis images at 100--150~MHz. Top row: the Cassiopeia~A field; bottom row: the near north celestial pole (NCP) calibration field. Left column: CLEAN images; right column: dirty images.}
  \label{fig:mfs_2x2}
\end{figure*}

\begin{figure*}
  \centering

  \begin{subfigure}[t]{0.49\textwidth}
    \centering
    \includegraphics[width=\linewidth]{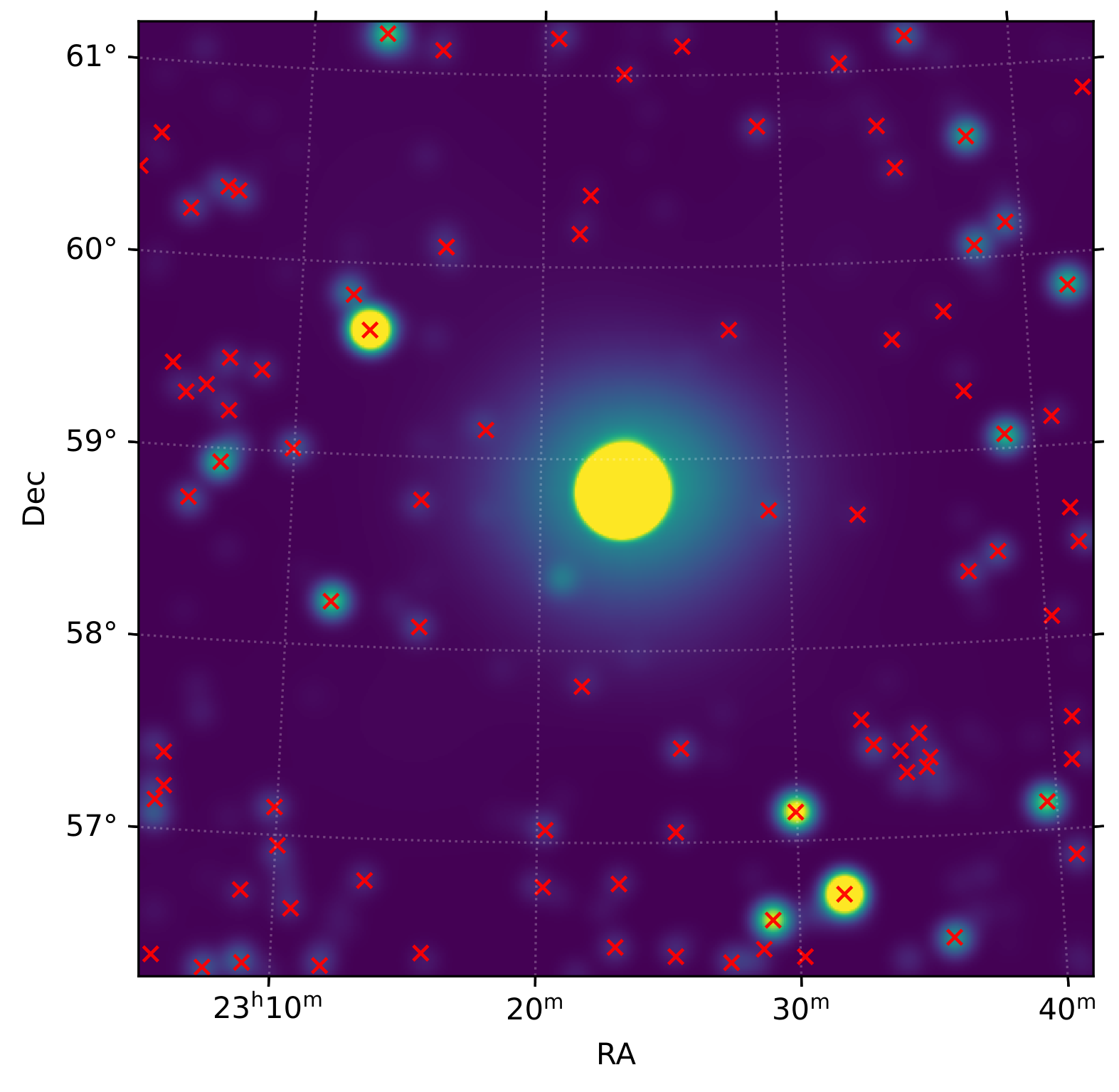}
    \caption{Cas~A: Sky model}
    \label{fig:casa_model}
  \end{subfigure}\hfill
  \begin{subfigure}[t]{0.49\textwidth}
    \centering
    \includegraphics[width=\linewidth]{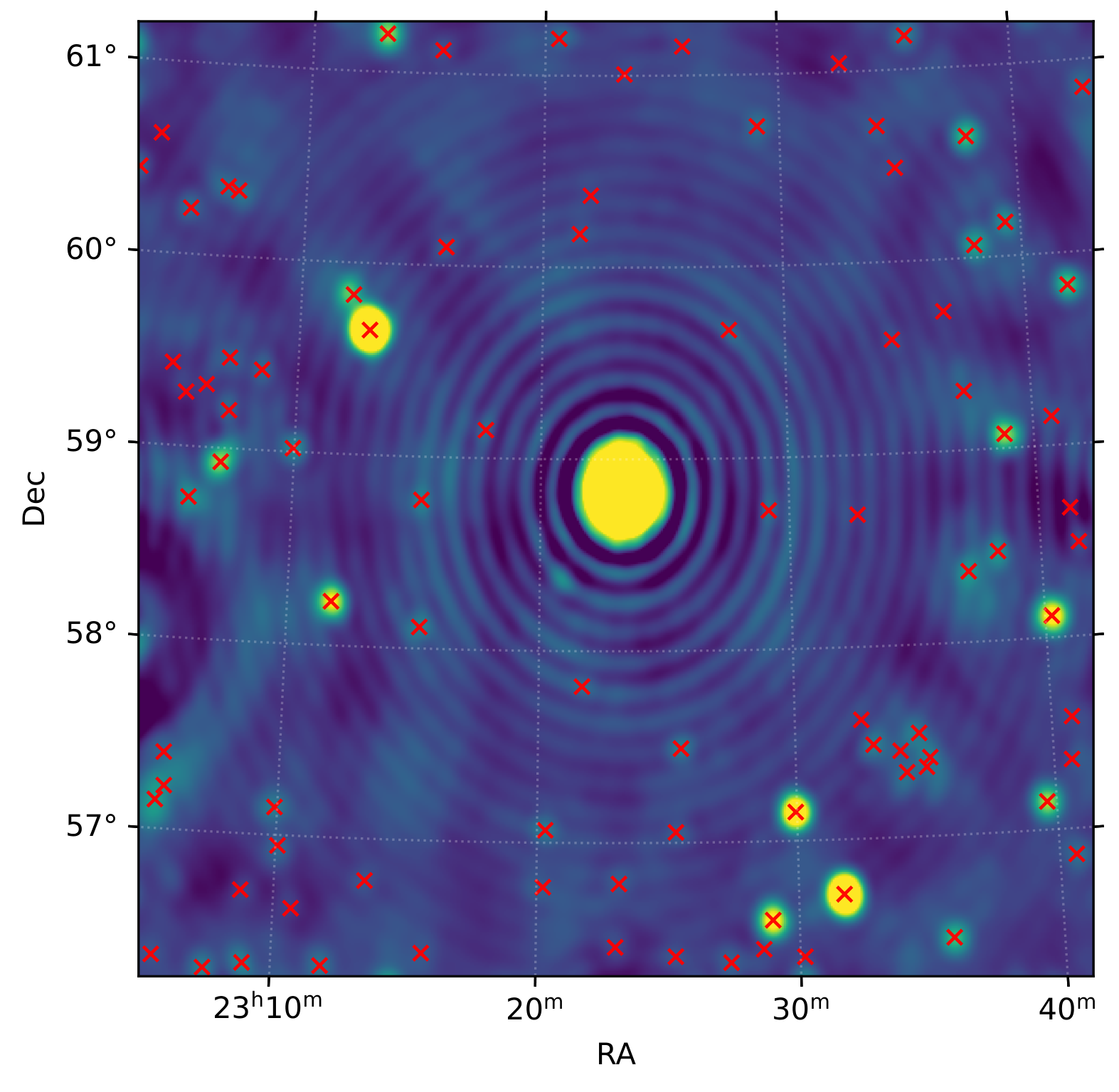}
    \caption{Cas~A: 100--150 MHz multi-frequency synthesis (MFS)}
    \label{fig:casa_data}
  \end{subfigure}

  \vspace{2mm}

  \begin{subfigure}[t]{0.49\textwidth}
    \centering
    \includegraphics[width=\linewidth]{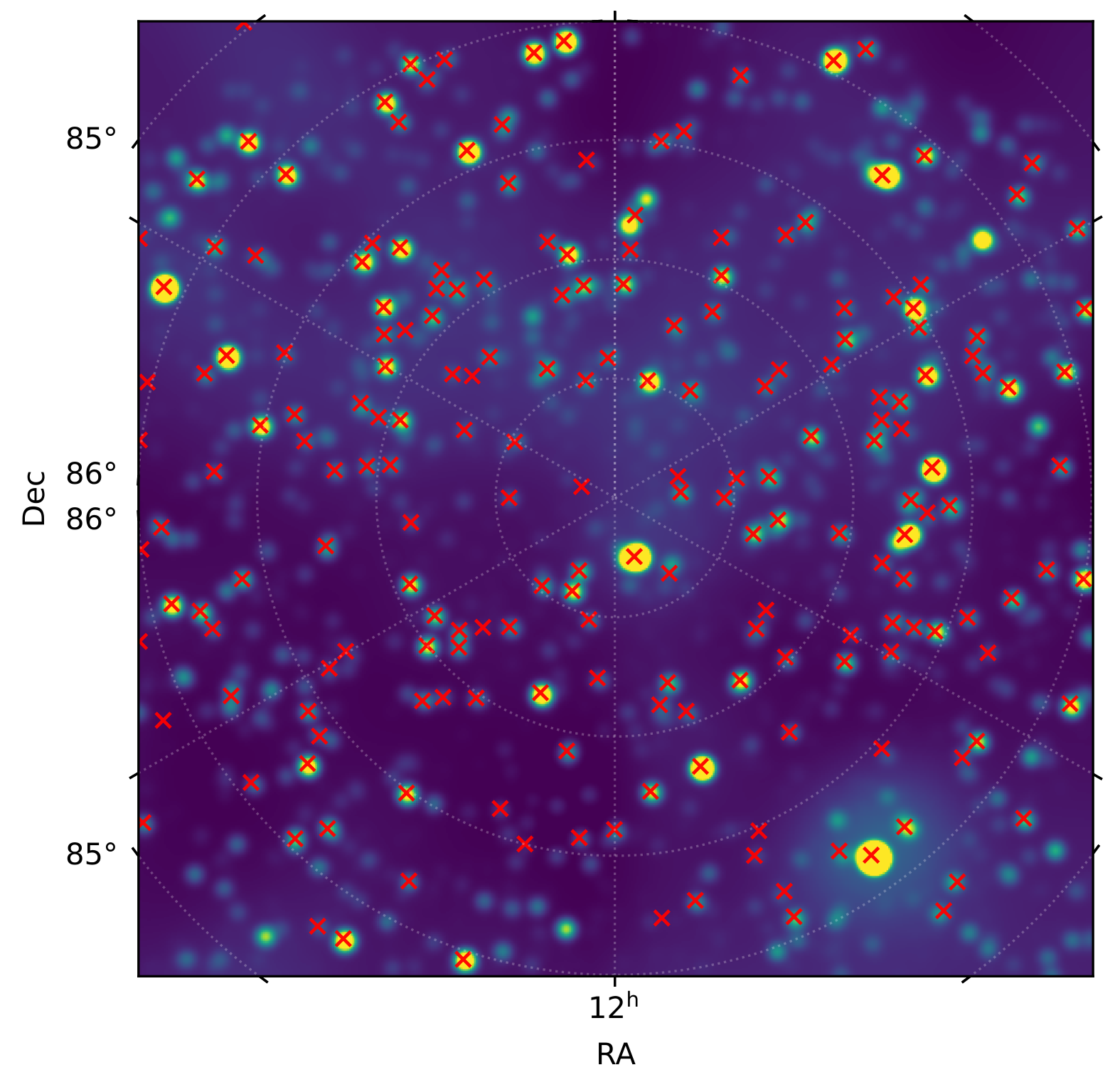}
    \caption{NCP: Sky model}
    \label{fig:ncp_model}
  \end{subfigure}\hfill
  \begin{subfigure}[t]{0.49\textwidth}
    \centering
    \includegraphics[width=\linewidth]{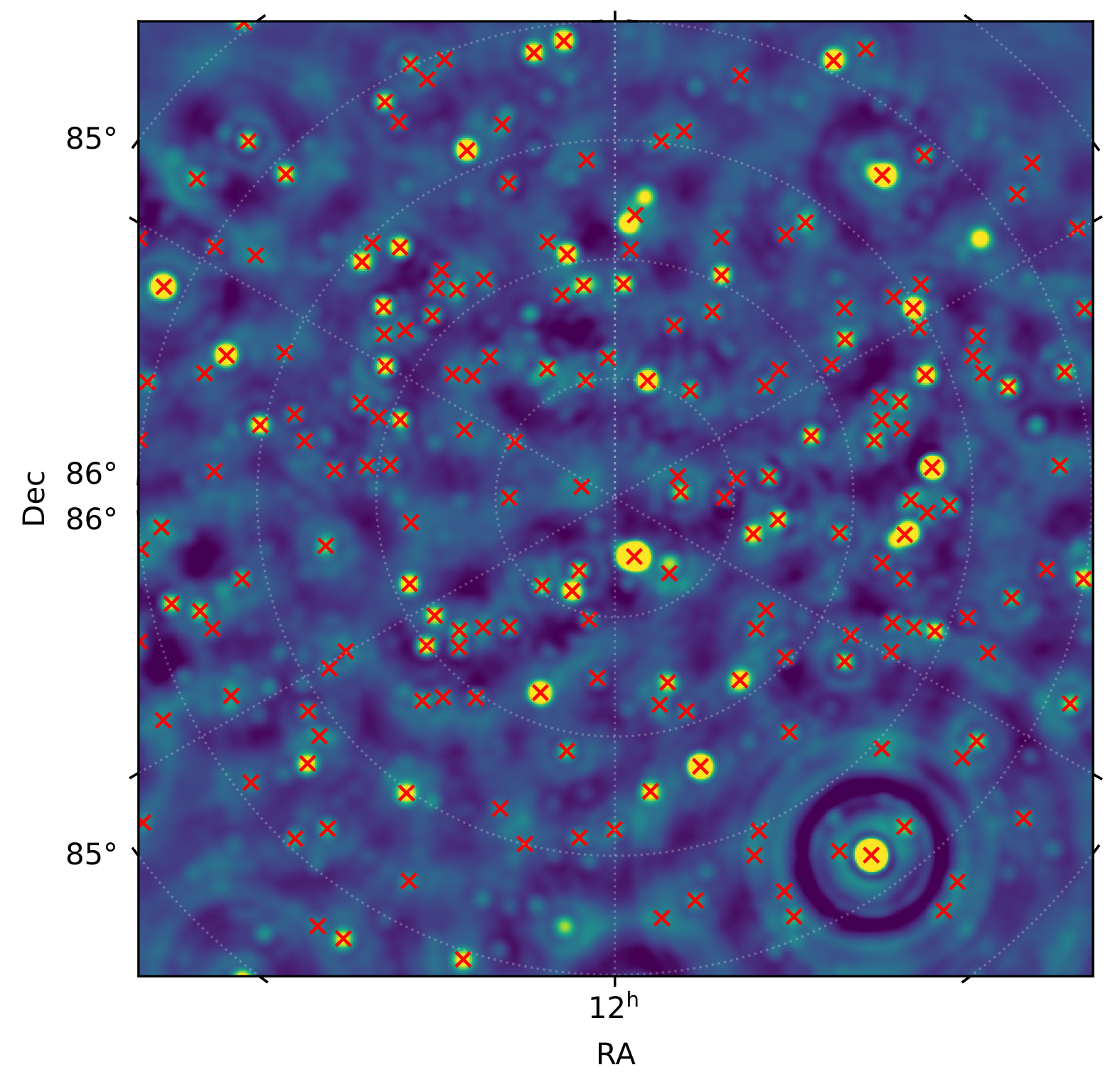}
    \caption{NCP: 100--150 MHz multi-frequency synthesis (MFS)}
    \label{fig:ncp_data}
  \end{subfigure}

  \caption{
Comparison of input sky models and simulated images.
(a,c) Input sky models (diffuse GSM model and point source model, convolved to the synthetic beam, reference frequency 125 MHz).
(b,d) Multi-frequency synthesis (MFS) images generated at 100--150 MHz based on simulated visibility.
Red crosses in the figure indicate point source locations identified from the MFS image.
  }
  \label{fig:model_data_compare_2x2}
\end{figure*}


\begin{figure}
  \centering

  \begin{subfigure}[t]{0.49\linewidth}
    \centering
    \includegraphics[width=\linewidth]{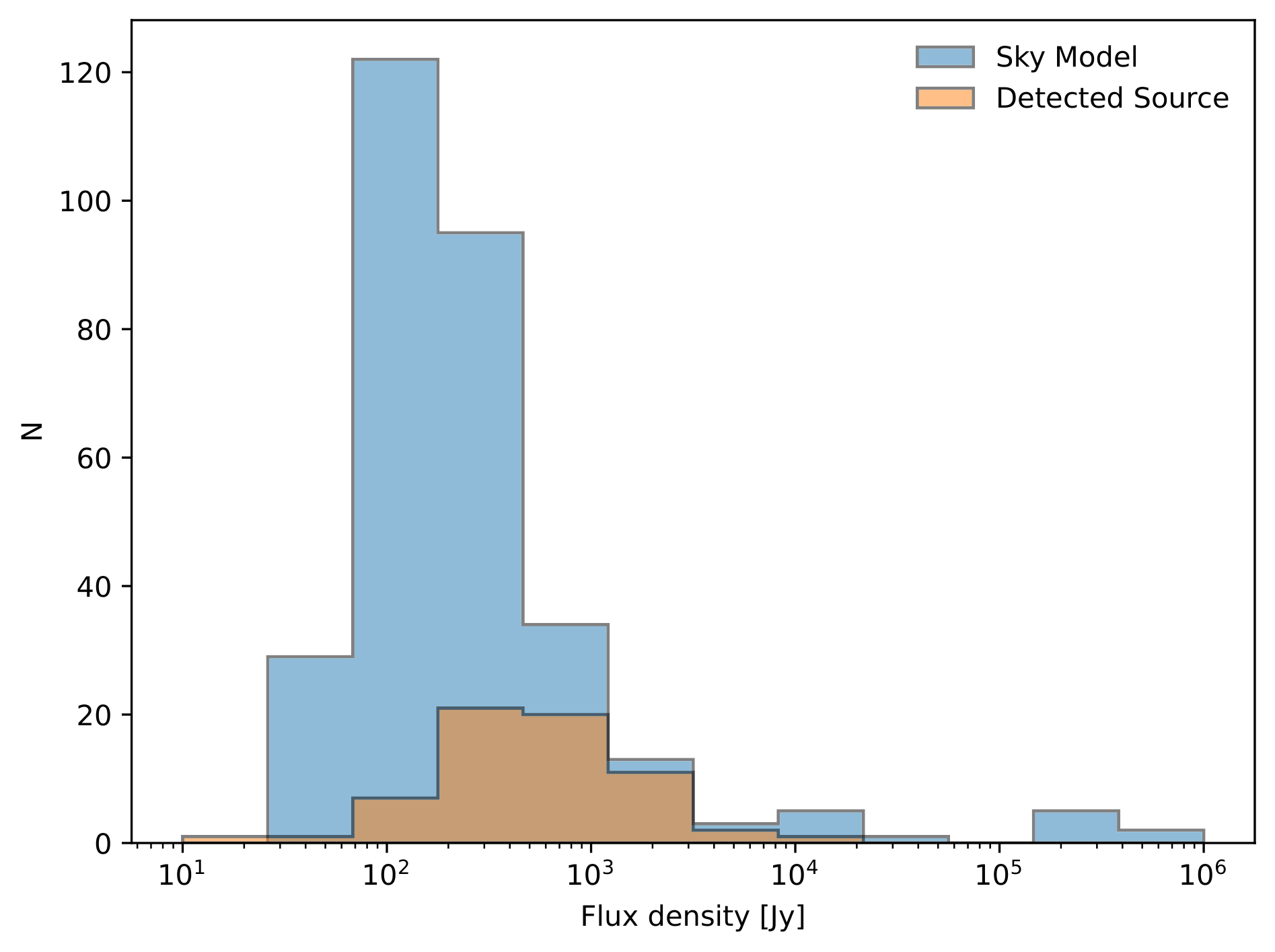}
    \caption{Cas~A field}
    \label{fig:fluxhist_casa_mfs_100_150}
  \end{subfigure}\hfill
  \begin{subfigure}[t]{0.49\linewidth}
    \centering
    \includegraphics[width=\linewidth]{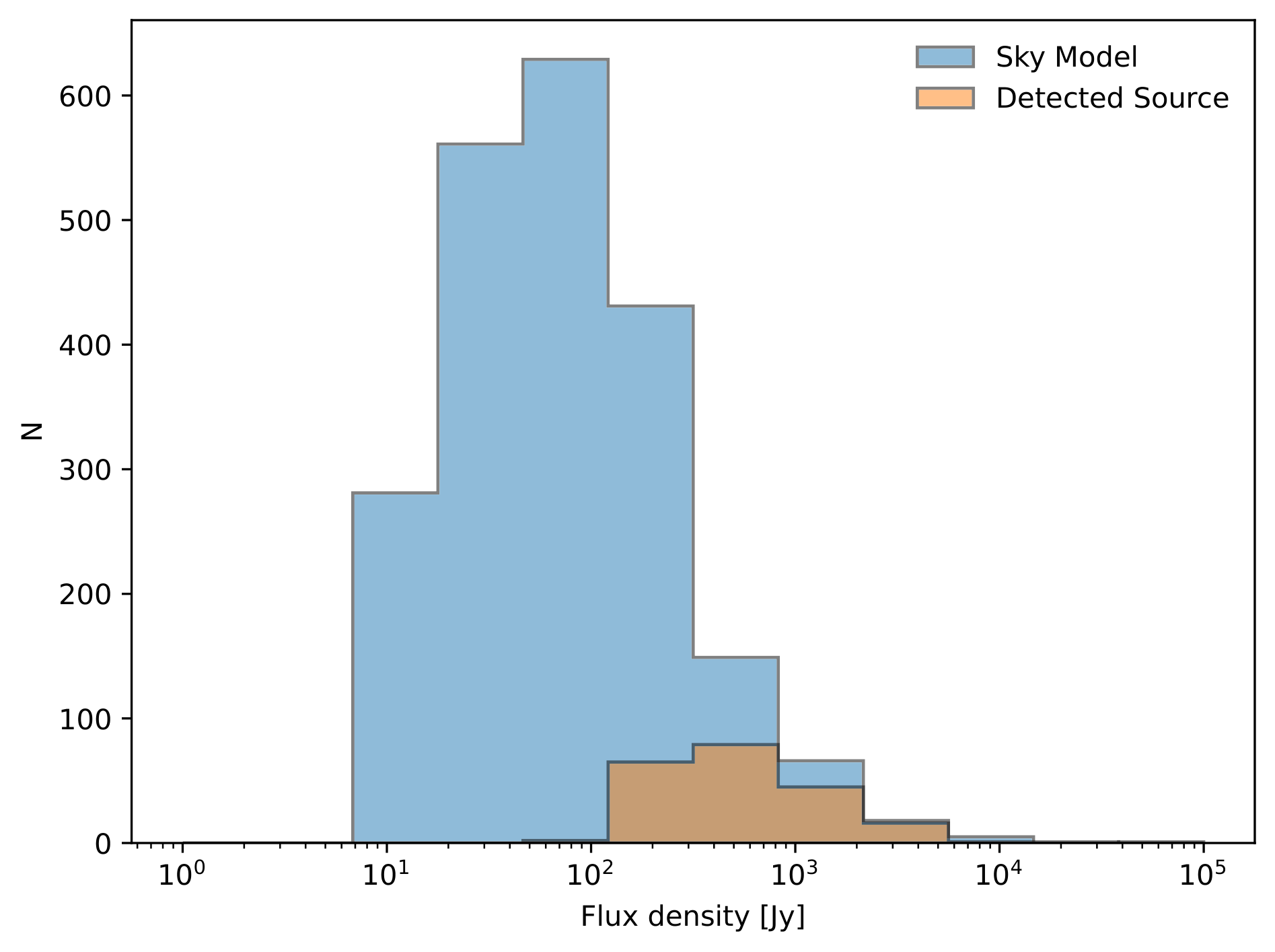}
    \caption{NCP field}
    \label{fig:fluxhist_ncp_mfs_100_150}
  \end{subfigure}

  \caption{
Comparison of the flux-density distributions between the input sky model (scaled to 125\,MHz) and the
sources detected from the 100--150\,MHz multi-frequency synthesis (MFS) images. The deficit of faint detections
is expected given the limited sensitivity and \emph{uv} coverage of the current four-station sub-array.
}
  \label{fig:fluxhist_mfs_100_150}
\end{figure}

\begin{table}
  \centering
  \caption{Background RMS ($\sigma_{\mathrm{bkg}}$) in the simulated multi-frequency synthesis CLEAN images.
  For each field, $\sigma_{\mathrm{bkg}}$ is measured in a selected low-fluctuation region on the 50--100~MHz image, and the same region is used for the other two bands.
  For reference, $\sigma_{\mathrm{th}}$ gives the expected thermal-noise level in the corresponding MFS image estimated from the injected visibility RMS. }
  \label{tab:mfs_rms}
  \begin{tabular}{cccccc}
    \hline
    Band (MHz) & $\nu_{\mathrm{c}}$ (MHz)
               & $\sigma_{\mathrm{th}}$ (Jy\,beam$^{-1}$)
               & $\sigma_{\mathrm{bkg}}^{\mathrm{NCP}}$ (Jy\,beam$^{-1}$)
               & $\sigma_{\mathrm{bkg}}^{\mathrm{Cas\,A}}$ (Jy\,beam$^{-1}$) \\
    \hline
     50--100   &  75  &  $2.14\times10^{-3}$  &  $7.61\times10^{1}$  &  $4.90\times10^{1}$ \\
     100--150  &  125 &  $6.18\times10^{-4}$  &  $6.19\times10^{1}$  &  $5.07\times10^{1}$  \\
     150--200  &  175 &  $2.91\times10^{-4}$  &  $1.67\times10^{1}$  &  $4.45\times10^{1}$  \\
    \hline
  \end{tabular}
\end{table}

In wide-field interferometric imaging, it is common to use multi-frequency synthesis (MFS) deconvolution together with $w$-projection or $w$-stacking to account for non-coplanar baselines. In the simulations presented here, we model a four-station east--west sub-array of the 21CMA (E13, E03, W02, and W09). The phase centers of the two simulated observations are located close to NCP. As a result, the associated w-term phase error can be estimated as $\Delta\phi_w \sim \pi w \theta^2$, where $w$ denotes the baseline component along the line of sight measured in units of the observing wavelength, and $\theta$ is the angular distance from the phase center in radians. For the configurations considered here, this phase error remains much smaller than unity. W-term effects can be safely neglected in the current version of the data-processing pipeline.

All simulated images presented in this work were produced with the WSClean imager using multi-frequency, multi-scale CLEAN. For the Cassiopeia~A field, we adopted a cell size of $35\arcsec$ and an image size of $512\times512$ pixels, corresponding to a field of view of about $5^\circ$. For the NCP field, we used a cell size of $56.25\arcsec$ with the same image size, giving an $8^\circ$ field of view. Deconvolution was performed with natural weighting and joint imaging of all 51 frequency channels, using up to 20\,000 CLEAN iterations, an automatic threshold of $3\sigma$, auto-masking at $5\sigma$, and a loop gain of 0.8. 
All images were made with WSClean using multi-frequency, multi-scale CLEAN on Stokes $I$ with natural weighting. We used a maximum of 20\,000 iterations, a fixed stopping threshold of 1.0~Jy\,beam$^{-1}$, and a loop gain of 0.8 (multiscale enabled; beam fitted). Dirty images were generated with identical settings but with $n_{\rm iter}=0$.
Within each 50\,MHz sub-band we adopt a single-term MFS model (no in-band spectral fitting), so the resulting MFS map is a frequency-weighted average referenced to the band-centre frequency $\nu_{\mathrm{c}}$. Therefore, no spectral index is assumed in the imaging step, and we do not attempt to recover explicit in-band spectral structure from these MFS images.
For diagnostic purposes, we also generated corresponding dirty images by repeating the imaging with identical settings but with $n_{\rm iter}=0$, which we use only to illustrate the synthesized beam and sidelobe structure. Representative CLEAN and dirty images for the Cassiopeia~A and NCP fields are shown in Figure~\ref{fig:mfs_2x2}. As an end-to-end sanity check, we directly compare the input sky models with the corresponding reconstructed images. For each field (Cas~A and NCP), we construct a reference sky-model image at 125\,MHz by combining the diffuse Global Sky Model (GSM) component with the point-source catalogue, and convolving the point-source component with the synthesized beam. We then identify point sources in the MFS images using PyBDSF (Python Blob Detector and Source Finder; \citealt{2015ascl.soft02007M}) and overplot their positions for reference. Figure~\ref{fig:model_data_compare_2x2} presents the comparison: panels (a,c) show the input sky models, and panels (b,d) show the reconstructed MFS images. The close agreement in source morphology and relative brightness provides a simple validation that the simulation-to-imaging chain reproduces the expected sky structure. As shown in Fig.~\ref{fig:fluxhist_mfs_100_150}, we also compare the flux-density distributions of the input sky model (scaled to 125\,MHz) and the sources detected from the 100--150\,MHz MFS images. Because the current experiment uses only four stations, the sensitivity and \emph{uv} coverage are limited; consequently, only a relatively small number of point sources are detected, and the effective point-source detection threshold is \(\sim 10^{2}\)~Jy.

To quantify background fluctuations, we measure the pixel RMS in the MFS CLEAN images for each field and frequency band. For each field, we define a single analysis region on the 50--100~MHz MFS image by sliding a window across the image and selecting the window that minimizes a robust RMS estimate. We compute the robust RMS by discarding the lowest and highest 5\% of pixel values within the window and taking the standard deviation of the remaining pixels. We then apply the same region to the other two bands of the same field to enable a consistent comparison across frequency. We refer to this metric as the background RMS, $\sigma_{\mathrm{bkg}}$. Table~\ref{tab:mfs_rms} summarizes the resulting $\sigma_{\mathrm{bkg}}$ values for the simulated MFS images of the NCP and Cassiopeia~A fields, together with the expected thermal-noise levels $\sigma_{\mathrm{th}}$ estimated from the injected visibility RMS. In these simulations, $\sigma_{\mathrm{bkg}} \gg \sigma_{\mathrm{th}}$, indicating that the measured background fluctuations are dominated by confusion and residual sidelobe structure rather than thermal noise.

After deconvolution, each beam image was divided by its corresponding primary-beam (PB) map to recover the intrinsic sky brightness distribution:
\begin{equation}
I_{\mathrm{pbcor}}(x, y) = 
\frac{I_{\mathrm{att}}(x, y)}{PB(x, y)} ,
\end{equation}
where $I_{\mathrm{att}}(x, y)$ is the attenuated image and $PB(x, y)$ is the primary-beam response at position $(x, y)$.

If different pointings overlap, the images were combined using PB-squared weighting:

\begin{equation}
I_{\mathrm{mosaic}}(x, y) = 
\frac{\displaystyle \sum_i PB_i^2(x, y)\, I_i(x, y)}
     {\displaystyle \sum_i PB_i^2(x, y)} ,
\end{equation}

which minimizes noise variations and ensures a uniform flux scale across the mosaic.

For beam-forming observations, the independently imaged beams were combined to form a final mosaic. Prior to mosaicking, each beam image was corrected for its frequency- and direction-dependent primary beam (PB) response. In the overlapping regions between beams, the data were weighted by the inverse variance of the local noise to ensure uniform sensitivity across the mosaic. For joint imaging, the calibrated Measurement Sets from all beams were supplied simultaneously to the imager, which performed deconvolution across the combined \emph{uv} coverage and applied PB corrections internally during gridding. In contrast, for linear mosaicking, individually deconvolved and PB-corrected beam images were regridded to a common astrometric frame and combined using the same noise-based weighting scheme. Consistency among beams was verified by comparing the flux density and astrometric alignment of sources in the overlap regions.  For the imaging of the two regions simulated in this work, mosaicking is not required. Therefore, the mosaicking stage is not fully implemented or tested in the present study. A complete implementation and validation of the final mosaicking stage will be explored in future work.

\begin{table}[htbp]
\centering
\caption{Summary of calibration and mosaicking workflow.}
\begin{tabular}{@{}lll@{}}
\toprule
\textbf{Stage} & \textbf{Objective} & \textbf{Implementation} \\ 
\midrule 
 RFI excision         & Remove corrupted visibilities     & Statistical 3$\sigma$ flagging and morphological filtering \\[3pt]
Per-beam calibration & Derive complex gains per beam     & Direction-independent and beam-aware solvers \\[3pt]
A-projection imaging & Apply PB response during gridding & Common Astronomy Software Applications (CASA), WSClean \\[3pt]
PB correction        & Restore intrinsic flux scale      & Division by PB or PB$^{2}$-weighted mosaicking \\[3pt]
Mosaic verification  & Ensure seamless beam transitions  & Overlap flux and position cross-checks \\
\bottomrule
\end{tabular}
\label{tab:workflow}
\end{table}

The workflow of data processing is shown in Table \ref{tab:workflow}. Overall, this pipeline reflects the contemporary calibration and imaging of the 21CMA observations. It enables wide-field low-frequency interferometry, particularly for instruments where beam variation, ionospheric phase structure, and extended emission recovery are critical to achieving high-fidelity science products.

\section{Discussion and Conclusions}
\label{sect:discussion}

In this work, we have developed and demonstrated a dedicated end-to-end simulation and data-processing framework for digital beamforming experiments using four stations of the 21 Centimeter Array (21CMA). The primary motivation is to quantify systematics, such as instrumental effects introduced by digital beamforming and two-stage channelization, and to provide a realistic test for validating calibration and imaging strategies prior to large-scale observations with the upgraded 21CMA system.

A key goal of this study is the construction of a set of simulations and pipelines that connect physically motivated sky models, station-level beamforming responses, realistic thermal noise, and full interferometric imaging. By adopting measured station layouts and a two-stage digital beamforming model consistent with the actual system architecture, the simulations capture chromatic and direction-dependent effects that are not present in idealized single-stage beamformers. In this work, each station is modeled as a beam-formed aperture array with identical isotropic, unpolarized elements, and incorporating realistic element patterns and full polarization response is deferred to future extensions. In particular, the analysis of two-stage channelization demonstrates that while on-axis sources are only weakly affected, off-axis sources exhibit characteristic piecewise-linear spectral modulations across coarse-channel boundaries. These features provide a clear diagnostic signature of two-stage digital beamforming and should be considered in high-dynamic-range imaging and spectral analyses.

The simulated observations of the Cassiopeia A field and a field near the North Celestial Pole demonstrate that the adopted calibration and imaging strategy is capable of recovering both compact and diffuse emission over a wide frequency range. Multi-frequency synthesis imaging combined with multi-scale deconvolution successfully reconstructs the dominant sky structures, and the measured background RMS levels are broadly consistent with expectations based on the injected thermal noise. We note that no RFI is injected in the simulations, so the results presented here correspond to the thermal noise case rather than the full RFI-mitigation conditions of real 21CMA data. The results show that current calibration approaches developed for traditional 21CMA observations can be adapted to digital beamforming modes, provided that direction-dependent effects and beam chromaticity are properly incorporated. The framework developed here allows systematic testing of calibration strategies, imaging parameters, and beam corrections under controlled conditions, and therefore serves as a valuable tool for pipeline development and performance verification.

Looking ahead, several extensions of this framework are both natural and necessary. Future work will incorporate more realistic antenna element patterns, full polarization response, explicit ionospheric phase-screen modeling, and time-variable beam perturbations, together with a complete treatment of multi-beam mosaicking for wide-field surveys. Beyond its immediate application to the four-station 21CMA beamforming experiment, the framework presented here has broader relevance for low-frequency aperture-array interferometers employing two-stage channelization and digital beam synthesis. The two-stage beamforming artifacts quantified in this work are generic to many modern station-based architectures, including those foreseen for SKA-Low. As such, the simulation and data-reduction pipeline provides a scalable and flexible platform for testing calibration robustness, spectral smoothness, and imaging fidelity under controlled conditions. This capability is particularly important for science cases that demand precise spectral characterization, including foreground mitigation for EoR/CD  studies, as well as other science projects that rely on stable, well-characterized station beams.

In summary, we have presented a comprehensive simulation and data-processing framework for four-station digital beamforming experiments with the 21CMA. This framework provides a unified pathway from synthetic visibilities to science-ready images and establishes a solid foundation for future improvements in beamforming data processing, calibration methodology, and system optimization for the upgraded 21CMA and related low-frequency aperture-array instruments.

\begin{acknowledgements}
This work was funded by the National SKA Program of China (Grant No. 2020SKA0110200). YY acknowledges the support of the Key Program of the National Natural Science Foundation of China (12433012).
\end{acknowledgements}

\bibliographystyle{raa}
\bibliography{bibtex}

\label{lastpage}

\end{document}